\documentclass[reprint,twocolumn,superscriptaddress,secnumarabic,amssymb, nobibnotes, aps, prl]{revtex4-1}

\usepackage{extarrows}
\usepackage{amsmath}    % need for subequations 
\usepackage{graphicx}   % need for figures
\usepackage{verbatim}   % useful for program listingsz
\usepackage{color}      % use if color is used in text
\usepackage{subfigure}  % use for side-by-side figures
\usepackage{hyperref}   % use for hypertext links, including those to external documents 
\usepackage{verbatim}   % and URLs
\usepackage[normalem]{ulem}

\begin{document}
	
\title{
%At the onset of thermal spin-state transition in LaCoO$_3$: \\
%Co $L_3$-edge RIXS and HB-DMFT perspective d
%Melting of excitonic dispersion in LaCoO$_3$: theory and experiment
Damping of spinful excitons in LaCoO$_3$ by thermal fluctuations: Theory and experiment
}

\author{Atsushi Hariki}
\thanks{These two authors contributed equally to this work.} 
\affiliation{Institute of Solid State Physics, TU Wien, 1040 Vienna, Austria}
\affiliation{Department of Physics and Electronics, Graduate School of Engineering, Osaka Prefecture University 1-1 Gakuen-cho, Nakaku, Sakai, Osaka 599-8531, Japan}

\author{Ru-Pan Wang}
\thanks{These two authors contributed equally to this work.}
\affiliation{Debye Institute for Nanomaterials Science, Utrecht University, Universiteitsweg 99, 3584 CG Utrecht, The Netherlands}

\author{Andrii Sotnikov}
\affiliation{Institute of Solid State Physics, TU Wien, 1040 Vienna, Austria}
\affiliation{Akhiezer Institute for Theoretical Physics, NSC KIPT, Akademichna 1, 61108 Kharkiv, Ukraine}
	
\author{Keisuke Tomiyasu}
\affiliation{Department of Physics, Tohoku University, Aoba, Sendai 980-8578, Japan}
\affiliation{NISSAN ARC, LTD., 1, Natsushima-cho, Yokosuka, Kanagawa 237-0061, Japan}
	
\author{Davide Betto}
\author{Nicholas B. Brookes}
\affiliation{European Synchrotron Radiation Facility, 71 Avenue des Martyrs, CS40220, F-38043 Grenoble Cedex 9, France}

\author{Yohei Uemura}
\affiliation{Debye Institute for Nanomaterials Science, Utrecht University, Universiteitsweg 99, 3584 CG Utrecht, The Netherlands}
%\affiliation{Paul Scherrer Institut, Forschungsstrasse 111, 5232 Villigen PSI, Switzer-lands}

\author{Mahnaz Ghiasi}
\affiliation{Debye Institute for Nanomaterials Science, Utrecht University, Universiteitsweg 99, 3584 CG Utrecht, The Netherlands}	

\author{Frank M. F. de Groot}
%\thanks{F.M.F.deGroot@uu.nl} 
\affiliation{Debye Institute for Nanomaterials Science, Utrecht University, Universiteitsweg 99, 3584 CG Utrecht, The Netherlands}
	
\author{Jan Kune\v s}
%\thanks{kunes@ifp.tuwien.ac.at}
\affiliation{Institute of Solid State Physics, TU Wien, 1040 Vienna, Austria}
\affiliation{Institute of Physics, Czech Academy of Sciences, Na Slovance 2, 182 21 Praha 8, Czechia}

\date{\today}
	
\begin{abstract}
We present Co $L_3$-edge resonant inelastic x-ray scattering (RIXS) of bulk LaCoO$_3$ across the thermally-induced spin-state crossover %($T=20,70,110,150$~K)
around 100~K. 
Owing to a high energy resolution of 25~meV, we observe unambiguously the dispersion of the intermediate-spin (IS) excitations in the low temperature regime.
Approaching the intermediate temperature regime, the IS excitations are damped and the bandwidth is reduced.
The observed behavior can be well described by a model of mobile IS excitons with
strong attractive interaction, which we solve using dynamical mean-field theory for hard-core bosons.
Our results provide a detailed mechanism of how high-spin (HS) and IS excitations
interact to establish the physical properties of cobaltite perovskites.
\end{abstract}
	
%%%%%%%%%%%%%%%%%%%%%%%%%%%%%%%%%%%%%%%%%%%%%%%%%%%%%
\maketitle

%%%%%%%%%%%%%%%%%%%%%%%%%%%%%%%%%%%%%%%%%%%%%%%%%%%%%
The physics of ionic insulators at energies below the band gap opened between fully occupied and empty atomic states is often trivial. Under certain conditions, electron-electron interaction may alter this picture by giving rise to low-energy bosonic excitations. This is the case of LaCoO$_3$, a structurally simple quasi-cubic material with complex magnetic and transport properties 
%that has been 
studied since 1950s~\cite{Goodenough1958jpcs,Heikes64,Raccah1967pr,Abbate1993prb,Asai1998jpsj,Stolen1997prb,Haverkort2006prl}.

LaCoO$_3$, a diamagnetic insulator with a low-spin (LS, $S=0$, $^1A_{1g}$) ground state and a band gap between filled $t_{2g}^6$ and empty $e_g^0$ subshells, becomes a paramagnetic insulator around 100~K.
This behavior is traditionally attributed to thermal population of excited atomic multiplets.
The high-spin (HS, $S=2$, $t_{2g}^4e_g^2$, $^5T_{2g}$) or intermediate-spin (IS, $S=1$, $t_{2g}^5e_g^1$, $^3T_{1g}$) 
nature of the lowest excited state has been the subject of an ongoing debate~\cite{Haverkort2006prl,Tanabe1954jpcj,deGroot1990prb,Zobel2002prb,Ishikawa2004prl,Korotin1996prb,Yan2004prb,Ropka2003prb,Podlesnyak2006prl,Noguchi2002prb,Maris2003prb,Vogt2003prb,Saitoh1997prb}. 
Both HS and IS scenarios evoke an important question.
Decorating the lattice with 
%a regular or random pattern of 
excited atoms leads to sizeable distribution Co-O bond lengths due to breathing distortion around HS atoms or Jahn-Teller distortion around IS atoms.
At experimentally reported concentrations, the excited atoms are expected to form a regular lattice, an effect favored by electron-lattice coupling~\cite{Bari72,Knizek09}, as well as 
%by 
electronic correlations~\cite{Kunes2011prl,Karolak15,Zhang12}.
Nevertheless, no spin-state order nor Co-O bond-length disproportionation was observed in LaCoO$_3$~\cite{Radaelli02}.
This leaves the possibility of dynamically fluctuating spin-state order~\cite{Doi2014prb}, for which, however, the picture of thermal atom-bound excitations provides no mechanism.

\begin{figure}[t] 
\includegraphics[width=1.00\columnwidth]{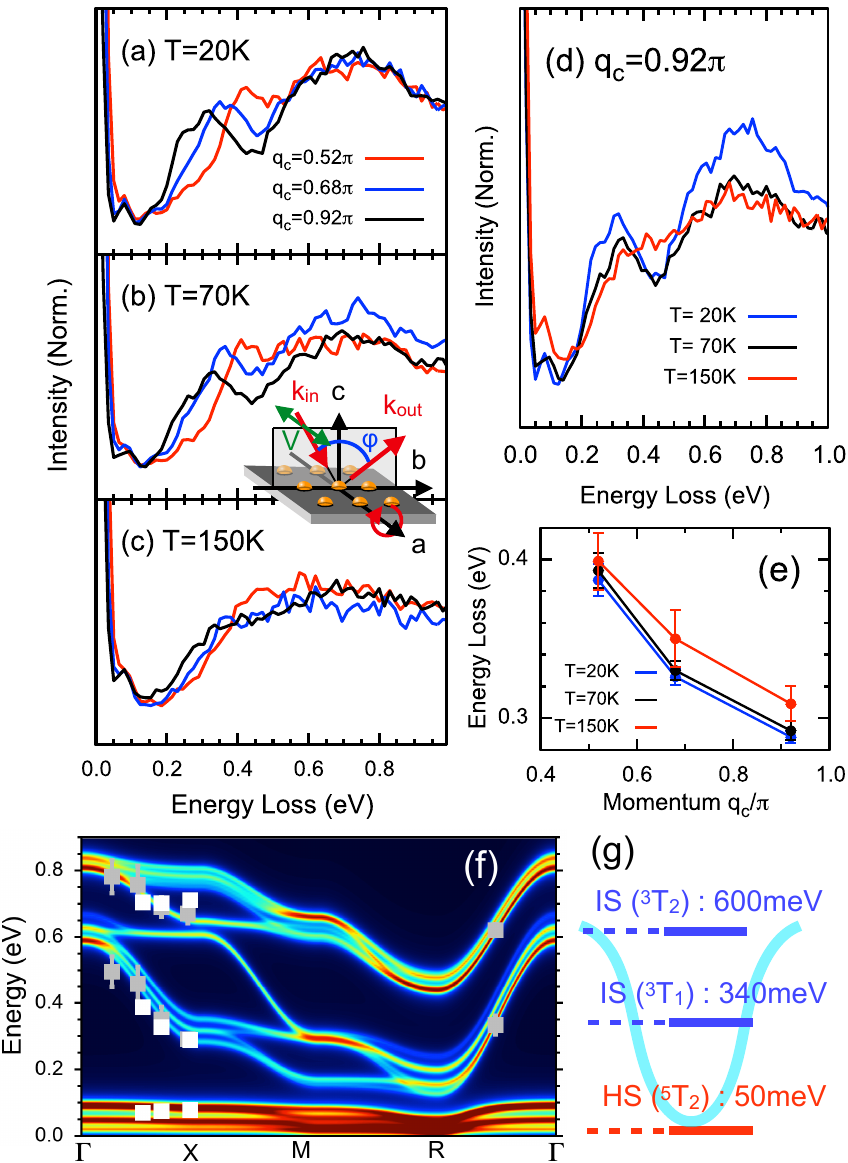}\hspace{0pt}
\caption{The experimental RIXS intensities for selected \textit{\textbf{q}}$=(0, 0, q_c)$ measured at (a) 20~K, (b) 70~K and (c) 150~K.
(d) temperature dependence at $q_c=0.92\pi$.
%%%%%%%%%%%%%%%%%%%%
Two distinct features located at 200--500~meV and 800~meV correspond to the $^3T_{1g}$ and $^3T_{2g}$ IS excitations~\cite{Wang2018prb}.
%%%%%%%%%%%%%%%%%%%%
The experimental geometry and the definition of angle $\varphi$ are illustrated in the inset.
(e) the peak position of the $^3T_{1g}$ IS excitation obtained by the fitting analysis, see SM~\cite{sm}.
(f) comparison at 20~K between theory (color map), present RIXS data with $\Delta E$=25~meV (white) and previous RIXS data with $\Delta E$=90~meV (gray)~\cite{Wang2018prb}. (g) sketch of the excitonic scenario:~the atomic-level energies and the IS dispersion on the LS background. 
}
\label{fig2}
\end{figure}

\begin{figure}[tb]
\includegraphics[width=0.95\columnwidth]{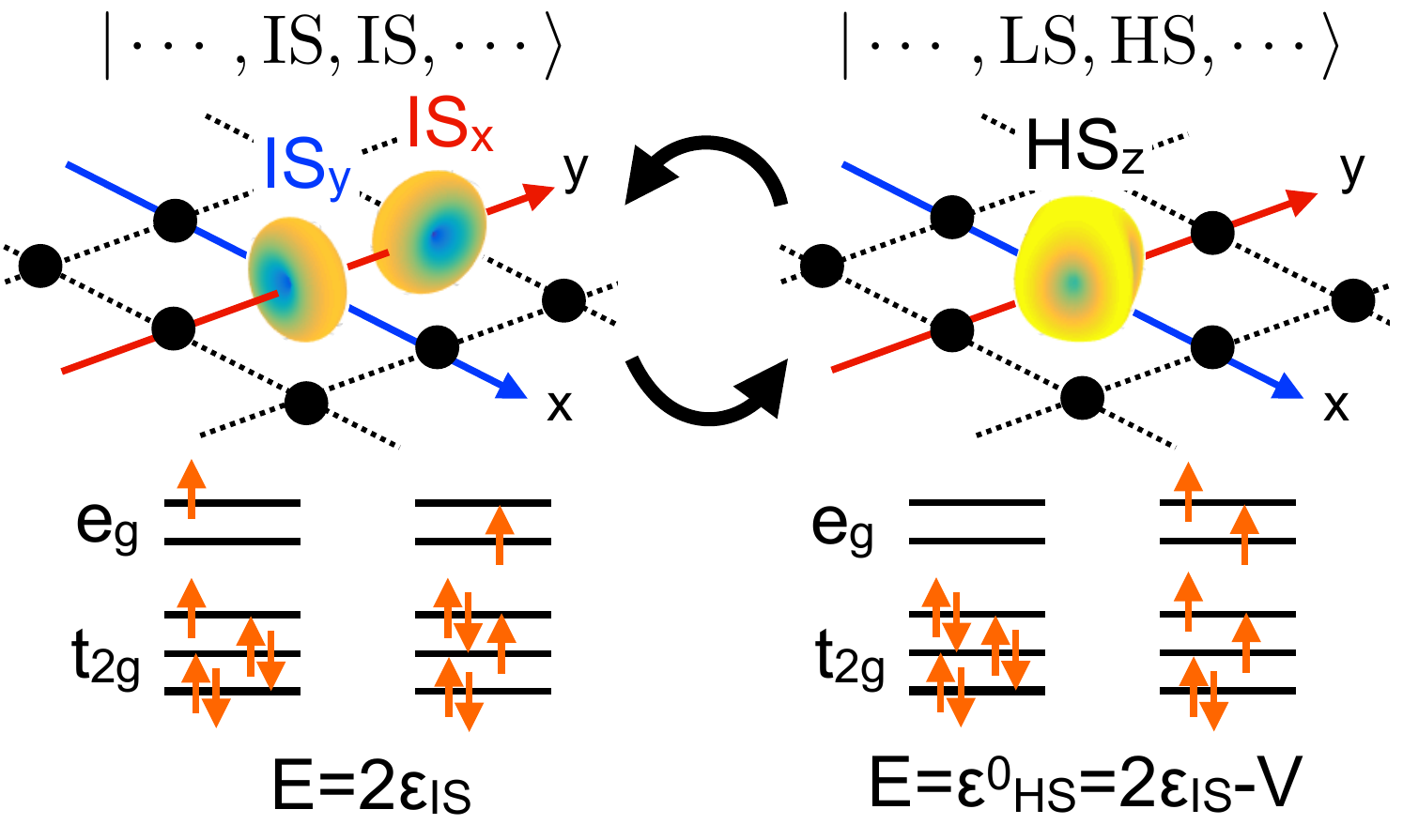}\hspace{0pt}
\caption{Sketch of the IS exciton (IS$_{x,y}$, ellipses correspond to the charge distribution) propagation on the LS (black circles) and the formation of the immobile HS bi-exciton (HS$_{z}$).
IS$_x$ (IS$_y$) excitons with the $d_{yz}\otimes d_{y^2-z^2}$ ($d_{zx}\otimes d_{z^2-x^2}$) character in the IS ($^3T_{1g}$) manifolds propagate in the $yz$ ($xz$) plane~\cite{Sotnikov2016sr,Afonso2017prb,Wang2018prb}.
IS$_z$ exciton (not shown) with the $d_{xy}\otimes d_{x^2-y^2}$ character propagates in the $xy$ plane.
The HS bi-exciton can be formed/melted by/into two IS excitons via the local attractive interaction between IS excitons.}
\label{fig1}
\end{figure}

Recently, a model of LaCoO$_3$ was proposed~\cite{Sotnikov2016sr},
%Sotnikov and Kune\v{s}~\cite{Sotnikov2016sr} proposed a model of LaCoO$_3$, 
in which the IS excitations are viewed as mobile bosons (excitons) carrying spin $S=1$, Fig.~\ref{fig2}g.
%Based on 
Band-structure calculations and experimental evidence lead to an estimated  excitonic half-bandwidth of 250~meV,
%to be 
comparable with the IS on-site energy of about $\epsilon_{\rm IS}=340$~meV.
The HS excitations are viewed as
%approximately
immobile $S=2$ bi-excitons 
%formed by IS excitons of different orbital flavors 
with 
%excitation 
energy of about $\epsilon_{\rm HS}=20$~meV~\cite{Podlesnyak2006prl,Tomiyasu2017prl}, see Fig.~\ref{fig1}.
This implies a 
%very 
strong 
%local 
attraction $V$ between IS excitons, $\epsilon_{\rm HS}\approx2\epsilon_{\rm IS}-V$.

RIXS experiments performed at low temperature matched nicely the generalized spin-wave theory based on first-principles parameters~\cite{Wang2018prb} and confirmed the estimate of Ref.~\onlinecite{Sotnikov2016sr}. The low-temperature RIXS measurements map out the dynamics of a single IS or HS excitation on the LS lattice (ground state), i.e., a single-boson problem.
The strong-coupling nature of the model~\cite{Sotnikov2016sr} suggests a sizeable temperature dependence of the excitation spectrum when excited states start being populated.
	
In this Letter, we report the thermal evolution of the IS dispersion in LaCoO$_3$ obtained with Co $L_3$-RIXS and its theoretical modelling using dynamical mean-field theory for hard-core bosons (HB-DMFT). 
Our main results are: i) observation of the low-temperature IS dispersion in the 0.2--0.5~eV range in the accessible part of the Brillouin zone with a high energy resolution, ii) observation of 
'melting' of the sharp IS dispersion into a narrow band of damped excitations at elevated temperatures, iii) theoretical modelling of the observed behavior by multi-flavor attractive Hubbard model for hard-core bosons.

%%%%%%%%%%%%%%%%%%%%%%%%%%%%%%%%%%%%%%%%%%%%%%%%%%%%%%%%%%
%%%%%%%%%%%%%%%%%%%%%%%%%%%%%%%%%%%%%%%%%%%%%%%%%%%%%%%%%%

{\it Experiment.} Co $L_3$-edge RIXS %($2p_{3/2}\rightarrow3d\rightarrow2p_{3/2}$)
was measured on the beamline ID32 at the European Synchrotron Radiation Facility (ESRF)~\cite{Brookes18}. The energy resolution $\Delta E$ was 25~meV. We recorded RIXS spectra at 20~K, 70~K, %110~K,
and 150~K. Note that 20~K (150~K) is well below (above) the spin-state crossover temperature 80--100~K.
The experimental geometry is illustrated in Fig.~\ref{fig2}. 
%The experimental geometry with is illustrated in Fig.~\ref{fig2}e.
The RIXS measurements were carried out in the $b$--$c$ scattering plane by rotating the sample around the $a$-axis, with linearly polarized x-rays (vertical to the $bc$-scattering plane).
%in the (pseudo) cubic axis of 
%with a lattice constant $a_{\rm cub}\approx$~3.83~\AA
The sample normal is aligned to the $c$-axis of the (pseudo) cubic crystal  with the lattice constant $a_{\rm cub}\approx$~3.83~\AA.
In this setup, we can measure a momentum transfer of  $\textit{\textbf{q}}=(0, 0, q_c)/a_{\rm cub}$.
%~\footnote{In the measurement, a small angle offset about 5$^\circ$ was applied to avoid a strong reflective signal. Its contribution $\delta$ as in $\textit{\textbf{q}}=(\delta, 0, q_c)$ is negligibly small ($\delta<$ 0.03 $\pi$)}. 
Hereafter, $a_{\rm cub}$ is omitted for simplicity. 
The x-ray wavelength at the Co $L_3$ edge in LaCoO$_3$ ($\approx15.9$~\AA) determines the accessible $|\textit{\textbf{q}}|$ values.
%The spectra were taken at four scattering angles $\varphi$ =~115$^\circ$, 103$^\circ$, 90$^\circ$, and 35$^\circ$, corresponding to \textit{\textbf{q}} = ($0, 0, 0.52\pi$), ($0, 0, 0.60\pi$), ($0, 0, 0.68\pi$), and ($0, 0, 0.92\pi$), respectively.
%\textcolor{red}{The momentum transfer is controlled by rotating the angle $\varphi$ in experimental geometry setting with an angle $\varphi$ as defined in the insert of Fig.~\ref{fig2}.
The spectra were taken at $\varphi$ =~115$^\circ$, 90$^\circ$, and 35$^\circ$, corresponding to \textit{\textbf{q}} = ($0, 0, 0.52\pi$), ($0, 0, 0.68\pi$), and ($0, 0, 0.92\pi$), respectively.
%
%
%
%
%
%At 20~K, $\varphi$=103$^\circ$, corresponding to \textit{\textbf{q}}=($0, 0, 0.60\pi$), is also measured.
%
%
%
%
%
%The RIXS spectra were fitted 6 Gaussian functions with respect to the excitons and one fluorescence profile, where the fluorescence profile is composed by 3 Gaussian functions and was used to normalize the spectra.
%After subtracting the fluorescence feature, the low-energy RIXS spectra were fitted with six Gaussians.
%The RIXS spectra were fitted with seven Gaussians and one fluorescence profile.
Details of the sample preparation and the data analysis including fitting can be found in Supplementary Material (SM).

Our main observations are summarized in Fig.~\ref{fig2}.
The RIXS spectra exhibit low-energy features observed in previous studies~\cite{Tomiyasu2017prl,Wang2018prb}. 
At 20~K, we observe a clear dispersion of the peak in 0.2--0.5~eV range that was assigned to the IS ($^3T_{1g}$) excitation~\cite{Tomiyasu2017prl,Wang2018prb}.  
%\textcolor{red}{In addition, the feature represents the HS excited state is clear observed.}
The HS ($^5T_{2g}$) excitation observed below $0.1$~eV shows no dispersion.
The IS dispersion is consistent with the theory and experiment of Ref.~\onlinecite{Wang2018prb}, see Fig.~\ref{fig2}f,
with substantially reduced error bars, provided by the present high energy resolution. 
Increasing the temperature above 100~K leads to a distinct narrowing and smearing of the dispersive feature, see Figs.~\ref{fig2}b-e. 
%\textcolor{red}{Put few words to discuss figure 1 e.}
%\textcolor{red}{One possibility will be the increase of the temperature induce the phonic broadening increase. But it can be rough estimated $\sim13$ meV for 150 K increase, where the broadening increase is much less than the separation of two IS states and will not induce the smearing of feature.}
%\textcolor{blue}{Note that the smearing cannot be ascribed to an overlap with another IS excitation ($^3T_{2g}$) by a possible broadening, such as phonon, that is expected to be much less than the separation between two IS ($\sim 300$~meV).}
%the observed smearing behavior (or the separation of two IS excitations ($^3T_{1g}$ and $^3T_{2g}$)).}
Other spectroscopic studies~\cite{Haverkort2006prl,Tomiyasu2017prl} indicated growing concentration of HS excitations with temperature, while the system remains a spatially uniform insulator.
Thus theoretical modelling of the heating effect on the dispersive IS excitations is a challenge that we address next.

%%%%%%%%%%%%%%%%%%%%%%%%%%%%%%%%%%%%%%%%%%%%%%%%%%%%%%%%%%
%%%%%%%%%%%%%%%%%%%%%%%%%%%%%%%%%%%%%%%%%%%%%%%%%%%%%%%%%%

{\it Theory.} 
An efficient description of
%In 
the insulating 
%state of 
LaCoO$_3$
%, i.e., the low and intermediate temperature regime, 
%the relevant Hilbert space may be
%\sout{reduced}
is provided by %derived
%by a strong-coupling 
%expansion,
%which leads to 
a low-energy effective model of the LS ground state %(vacuum) 
and its bosonic IS and HS excitations. Such a model was constructed in Ref.~\onlinecite{Wang2018prb} starting from density-functional theory. 
Despite substantial simplification, the bosonic model 
%still 
poses a 
%complicated 
strongly interacting problem, except for the lowest temperatures $T\approx 0$, where thermal excitations vanish and the interactions between them does not matter.
The RIXS final state at $T\approx 0$ 
%limit 
is a single IS (or HS) state excited on %an otherwise 
the LS lattice, a %non-interacting 
state that can be described with a generalized linear spin-wave theory~\cite{Wang2018prb}, 
see Fig.~\ref{fig2}f.

%The final state of RIXS process a single IS (or HS) state in otherwise LS lattice, a non-interacting state that can be captured with a generalized linear spin-wave theory, as in Ref.~\onlinecite{Wang2018prb} reproduced in Fig.~\ref{fig1}g.

%as described in Ref.~\onlinecite{Wang2018prb}. It proceeds in following steps. After a density functional calculation, a Hubbard model spanning the Co-$d$ bands is constructed~\cite{wien2k,wien2wannier,wannier90}. The next step is the Schrieffer-Wolff projection~\cite{Schrieffer66} on the subspace containing the lowest Co $3d^6$ multiplets.
%Keeping only the terms that describe creation/annihilation of an atomic excitation out of the LS ground state, we arrive at a non-interacting bosonic Hamiltonian, which describes dynamics of a single excitation relevant at low temperature.

%The hallmark of LaCoO$_3$ is the strong temperature dependence of physical properties. 
At elevated temperatures, the strongly interacting excitations must be taken into account. A direct treatment of the model
of Ref.~\onlinecite{Wang2018prb} with 24 states per each Co atom is technically hopeless and %would 
obscuring the key physics. Therefore we simplify the model to the essential minimum:
i) we neglect the spin structure of the problem
%\textcolor{blue}{(i.e.,~internal spin-triplet structure in $^3 T_{1g}$ IS)},
ii) we neglect the spin-orbit coupling,
iii) we approximate the HS excitation as two IS excitations on the same site,
iv) we use the DMFT approximation~\cite{Georges96,Byczuk08}.
%to describe the strong interaction accurately.
%We keep the three orbital flavors of IS as well as HS excitations.
%However, as we show in SM~\cite{sm}, 
%the same qualitative behavior is obtained with two IS and one HS flavor.
Our %simplified 
model has the form of three-flavor bosonic Hubbard Hamiltonian
on a cubic lattice
%%%%%%%%%%%%%%%%%%%%%%%%%%%%%%%%%%%%
%\begin{eqnarray}
%\label{eq:model}
%\hat{H} &=& \epsilon_{\rm IS}\sum_{\bold{i}}\sum_{\gamma}\hat{n}_{\bold{i},\gamma}
%-V\sum_{\bold{i}}\sum_{\langle \gamma,\gamma' \rangle}
%\hat{n}_{\bold{i},\gamma}\hat{n}_{\bold{i},\gamma'} \notag \\
%&+&t\sum_{\bold{i}}\sum_{\gamma}\sum_{\bold{e_\gamma}}
%(\hat{b}^\dag_{\bold{i}\pm\bold{e_\gamma},\gamma} %\hat{b}^{\phantom\dag}_{\bold{i},\gamma}
%+\hat{b}^\dag_{\bold{i}\pm\bold{e}_z,\gamma} %\hat{b}^{\phantom\dag}_{\bold{i},\gamma}).
%\end{eqnarray}
%%%%%%%%%%%%%%%%%%%%%%%%%%%%%%%%%%%%
\begin{eqnarray}
\label{eq:model}
\hat{H} &=& \epsilon_{\rm IS}\sum_{\bold{i}}\sum_{\alpha}\hat{n}_{\bold{i},\alpha}
-V\sum_{\bold{i}}\sum_{\alpha>\beta}
\hat{n}_{\bold{i},\alpha}\hat{n}_{\bold{i},\beta} \notag \\
&+&t\sum_{\bold{i}}\sum_{\alpha,\beta}
(1-\delta_{\alpha\beta} )\hat{b}^\dag_{\bold{i}\pm\bold{e}_{\beta},\alpha} \hat{b}^{\phantom\dag}_{\bold{i},\alpha}.
\end{eqnarray}
%%%%%%%%%%%%%%%%%%%%%%%%%%%%%%%%%%%%
The vacuum of the model represents the global LS state
$|\emptyset\rangle=\prod_\bold{i}|\rm{LS}_\bold{i}\rangle$.
Bosonic $\hat{b}^\dag_{\bold{i},\gamma}$ operators create the IS excitations 
$|{\rm IS}_{\bold{i},\gamma}\rangle\equiv
\hat{b}^\dag_{\bold{i},\gamma}|\emptyset\rangle$
with the orbital symmetry
$\varepsilon_{\alpha\beta\gamma}d_{\alpha\beta}\otimes d_{\alpha^2-\beta^2}$, with $\alpha$, $\beta$, $\gamma$ from $\{x,y,z\}$ and $\varepsilon_{\alpha\beta\gamma}$ being the fully anti-symmetric tensor. HS excitations are represented by a doubly occupied sites
$|{\rm HS}_{\bold{i},\gamma}\rangle\equiv\|\varepsilon_{\alpha\beta\gamma}\|\hat{b}^\dag_{\bold{i},\alpha}\hat{b}^\dag_{\bold{i},\beta}|\emptyset\rangle$. The Pauli principle of the underlying fermionic problems leads to kinematic hard-core constraints of maximum one boson of a given flavor and a maximum total of two bosons 
per site. 
The constraints can be enforced explicitly or dynamically by introducing an additional on-site interaction $U\sum_\bold{i}(\sum_\alpha (1-\hat{n}_{\bold{i}\alpha})\hat{n}_{\bold{i}\alpha}
+\hat{n}_{\bold{i}x}\hat{n}_{\bold{i}y}\hat{n}_{\bold{i}z})$ with
$U\rightarrow\infty$. The $b$-excitations can propagate by hopping to nearest-neighbor
sites in direction $\bold{e}_{\alpha}$. Due to the shape of the $T_{1g}$ excitations, the hopping
of IS$_{\gamma}$ along $\gamma$-axis is negligible compared
to the two perpendicular directions. 

To summarize, the model describes a gas of mobile bosons
of three flavors
($T_{1g}$ IS)
that interact via local attractive (inter-flavor) interaction.
%\textcolor{blue}{The HS ($T_{2g}$) state can be formed by the two IS bosons, thus regarded as bi-exciton, and it can be decomposed into the two IS}.
The LS $\rightarrow$ IS transition at an x-ray excited Co atom in RIXS corresponds to creating a boson in (\ref{eq:model}).
The $T_{1g}$ IS-part of
the RIXS spectra therefore corresponds to
the positive frequency part of the single particle spectral function of the $b$-bosons in the model.

We use the following model parameters $t=58$~meV, $\epsilon_{\rm{IS}}=340$~meV, $V=620$~meV. The values of $\epsilon_{\rm{IS}}$ and $t$
were obtained in Ref.~\onlinecite{Wang2018prb} by
a combination of first-principles calculation and RIXS analysis.
Nevertheless, their magnitudes can be estimated using the
%\sout{typical}
established
values of crystal-field splitting, Hund's exchange and $e_g$ ($t_{2g}$) bandwidths in
%\sout{$3d$ transition-metal perovskites}
LaCoO$_3$
together with the experimental fact of stability of the LS ground state and
%\sout{existence}
activation energies
of low-lying 
%\sout{(15-20~meV)}
excited states.
The present $\epsilon_{\rm{IS}}$ compares well to
the estimate provided by ligand-field analysis~\cite{Haverkort2006prl}.
The value of $t$ reflects the hopping amplitudes (bandwidths) of the $t_{2g}$ and $e_g$ electrons in LaCoO$_3$~\cite{Sotnikov2016sr}, 
while it depends to some extent on the effective Hubbard repulsion within the Co $3d$ shell
(determined by matching to the RIXS experiment~\cite{Wang2018prb}).
The value of $V$ is strongly constrained by the basic fact of stability of the
LS ground state and existence of HS excitations at 15-20~meV~\cite{Haverkort2006prl,Podlesnyak2006prl}.
This implies $0\lessapprox \epsilon^0_{\rm{HS}}=2\epsilon_{\rm{IS}}-V$. While it is the interaction $V$
that appears in our theory, we will use the more natural parameter $\epsilon^0_{\rm{HS}}$
in the following discussion, although it is not directly experimentally observable as will be become clear later.

We treat model (\ref{eq:model}) using the bosonic DMFT~\cite{Byczuk08,Anders10,Anders11,Hu09} with strong-coupling continuous-time quantum Monte-Carlo (CT-QMC) impurity solver~\cite{Anders11,Werner06}. 
%in a parameter range without condensate.
The hard-core constraint can be implemented dynamically by introducing a large intra-flavor interaction or explicitly~\footnote{We have checked on several cases that both approaches lead to the same low-energy spectral functions and excitation concentrations, see SM~\cite{sm}.}.
Nevertheless, the explicit constraint has numerical advantages and is more elegant.
While enforcing the constraint in the CT-QMC calculation is straightforward, care must be taken with the definition of self-energy, because the  Green's function does not have the canonical $\tfrac{1}{i\omega_n}$ high frequency limit due to the constraint.
%, where $\omega_n=2\pi n/\beta$ is the bosonic Matsubara frequency with the inverse temperature $\beta=1/k_BT$.
An analogous problem arising in the $t-J$ model for fermions was discussed by Shastry~\cite{Shastry11}
and Perepelitsky and Shastry~\cite{Perepelitsky13}, who introduced the concept of extremely correlated Fermi liquid and  a modified Dyson equation, which for bosons takes the form
%%%%%%%%%%%%%%%%%%%%%%%
%\begin{equation}
%    G_k(i\omega_n)=\frac{1-2 \langle \hat{n}\rangle }{i\omega_n-(1-2\langle \hat{n}\rangle
%    )\epsilon_k-\Sigma(i\omega_n)}.
%\end{equation}
%%%%%%%%%%%%%%%%%%%%%%%
%%%%%%%%%%%%%%%%%%%%%%%
\begin{equation}
    G_{\mathbf{k}}(i\omega_n)=\frac{1-\eta}{i\omega_n-(1-\eta)\epsilon_{\mathbf{k},\gamma}-\Sigma(i\omega_n)},
\end{equation}
%%%%%%%%%%%%%%%%%%%%%%%
where $\eta$ measures the spectral weights missing due to the hard-core constraint and is given as
$\eta=2\langle \hat{n}_\alpha\rangle+
2\langle \hat{n}_\alpha\hat{n}_\beta \rangle_{\alpha\neq\beta}$, where the occupations are equal for all flavors.
We refer the reader interested in technical details to SM~\cite{sm} and focus on discussion of the results.
\begin{comment}
where the spectral weight in the numerator depends on the flavor occupation $\hat{n}$ and thus differs from 1, while the self-energy $\Sigma(i\omega_n)$ is local and has the standard ${\Sigma(i\omega_n\rightarrow\infty)\sim\Sigma'+\tfrac{\Sigma''}{i\omega_n}}$ high-frequency behavior.
Note that the numerator in the case of hard-core bosons differs from the fermionic case and can go negative. The Green's function in the present case is diagonal in the bosonic flavor and $\epsilon_{k}=\epsilon_{\rm{IS}}+2t\left(\cos(k_z)+\cos(k_\gamma)\right)$, where $k_\gamma=k_x,k_y$ depending on the flavor. The DMFT self-consistent loop is closed by calculating the hybridization function (Weiss field) of the auxiliary impurity problem
%%%%%%%%%%%%%%%%%%%%%%%%%%%%%%%%%%%%%%%%%%%%%%%%%%%%%%%%%%%%%%%%%%%%%%%%%%%%%%%%%%
\begin{equation}
    \Delta(i\omega_n)=\frac{i\omega_n-\Sigma(i\omega_n)}{1-2\langle \hat{n}\rangle}
    -\epsilon_{\rm{IS}}-G_{\rm{loc}}^{-1}(i\omega_n),
\end{equation}
%%%%%%%%%%%%%%%%%%%%%%%%%%%%%%%%%%%%%%%%%%%%%%%%%%%%%%%%%%%%%%%%%%%%%%%%%%%%%%%%%%
where $G_{\rm{loc}}=\sum_k G_k$ is the local Green's function, which is the same for both bosonic flavors in our model.
\end{comment}

%%%%%%%%%%%%%%%%%%%%%%%%%%%%%%%%%%%%%%%%%%%%%%%%%%%%%%%%%%
\begin{figure}[t]
\includegraphics[width=0.92\columnwidth]{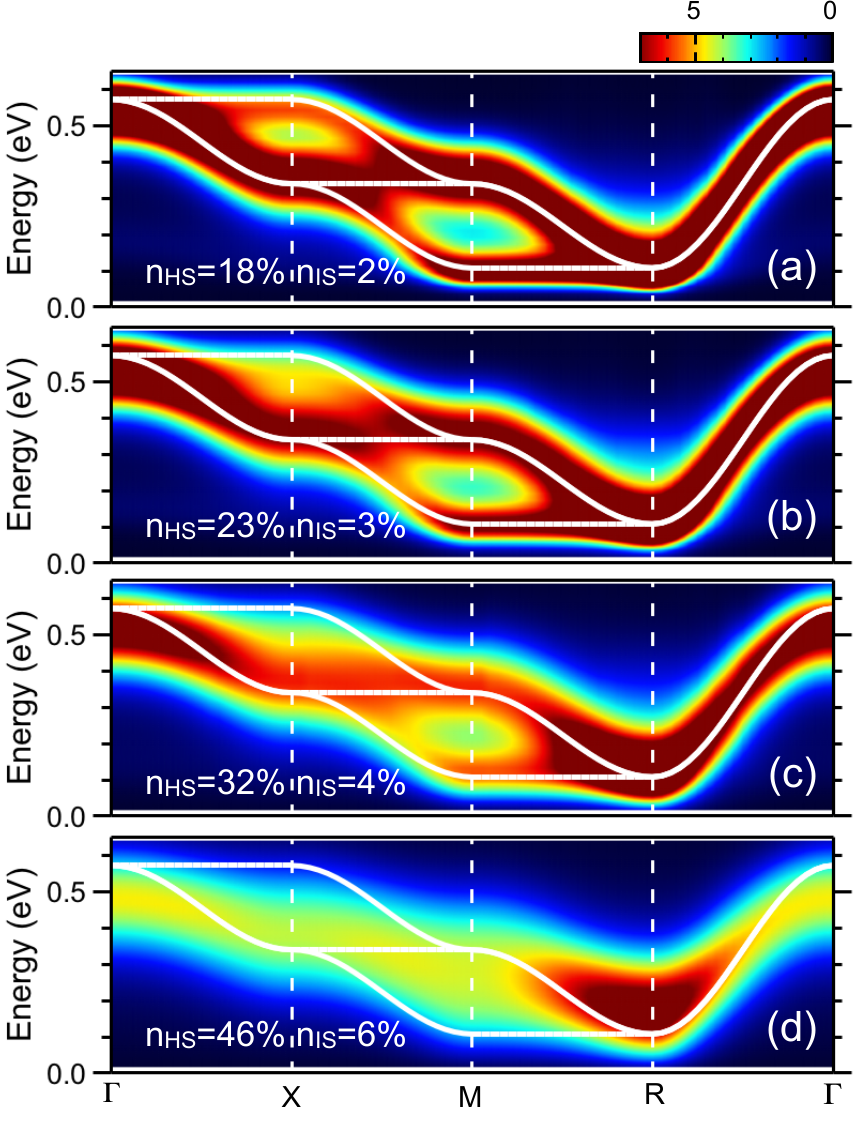}\hspace{0pt}
\caption{The $k$-resolved excitation spectra of the attractive Hubbard model obtained by HB-DMFT for various HS and IS populations at $V=620$~meV and (a) $\beta=80$, (b) $\beta=60$, (c) $\beta=40$, and (d) $\beta=20$.
Here $\beta$ is the inverse temperature $\beta=1/k_{B}T$.
The white line indicates the non-interacting dispersion at $0$~K.
}
\label{fig3}
\end{figure}
%%%%%%%%%%%%%%%%%%%%%%%%%%%%%%%%%%%%%%%%%%%%%%%%%%%%%%%%%%

%The calculations were performed for $t=58$~meV ($W_{\rm IS}=468$~meV), $\epsilon_{\rm{IS}}=340$~meV, following the estimation above, see Fig.~\ref{fig2}d and SM~\cite{sm}.
%We have used several values of the on-site attraction $V$ (the results are summarized in SM~\cite{sm}), out of which we picked $V=620$~meV.
Figs.~\ref{fig3}a--d show the evolution of the positive frequency part of $k$-resolved spectral function $A_{\mathbf{k}}(\omega)=-\tfrac{1}{\pi}\operatorname{Im}G_{\mathbf{k}}(\omega^+)$ with temperature $T$.
Despite its simplicity, the model (\ref{eq:model}) captures the essential features of the experimental data (Figs.~\ref{fig2}a--d). 
At $T=0$ the system is in its vacuum ground state (all atoms in the LS state) and the spectrum consists of three bands with $\epsilon_{\mathbf{k},\gamma}=\epsilon_{\rm{IS}}+2t\sum_{\alpha} (1-\delta_{\alpha\gamma})\cos k_{\alpha}$ dispersion indicated by white lines. This dispersion, by construction, reproduces the $T_{1g}$ part of the spectrum in Fig.~\ref{fig2}f except for the effect of the spin-orbit coupling neglected in model (\ref{eq:model}). 
The HB-DMFT spectrum in Fig.~\ref{fig3}a taken at slightly elevated temperature closely resembles the $T=0$ spectrum. 
Heating the system further, the equilibrium state changes due thermal
population
of the excited states. These are dominantly  HS excitations, 
the doubly occupied sites
$n_{\rm{HS}}=\langle \hat{n}_x \hat{n}_y\rangle + \langle\hat{n}_y \hat{n}_z\rangle + \langle\hat{n}_z \hat{n}_x\rangle$.
The IS population, i.e., the concentration of singly occupied sites
$n_{\rm{IS}}=\langle \hat{n}_x\rangle+\langle \hat{n}_y\rangle  +\langle \hat{n}_z\rangle -2n_{\rm{HS}}$, is
9 to 7 times smaller.
Increasing concentration of the excitations results in band narrowing, broadening of the spectral lines and spectral weight redistribution, Figs.~\ref{fig3}cd.

When accessing the role of the attractive interaction $V$, one has to consider that even a model with $V=0$ exhibits some $T$-dependence of the spectrum due to the kinematic hard-core constrains.
However, the thermal population of the excited states 
for $V=0$ remains small in the studied temperature range and the $T$-dependence of the spectra is negligible.
Formation of HS excitations, represented by strong attractive $V$, is thus crucial for the observed behavior of the RIXS spectra.

\begin{figure}[t]
\includegraphics[width=1.00\columnwidth]{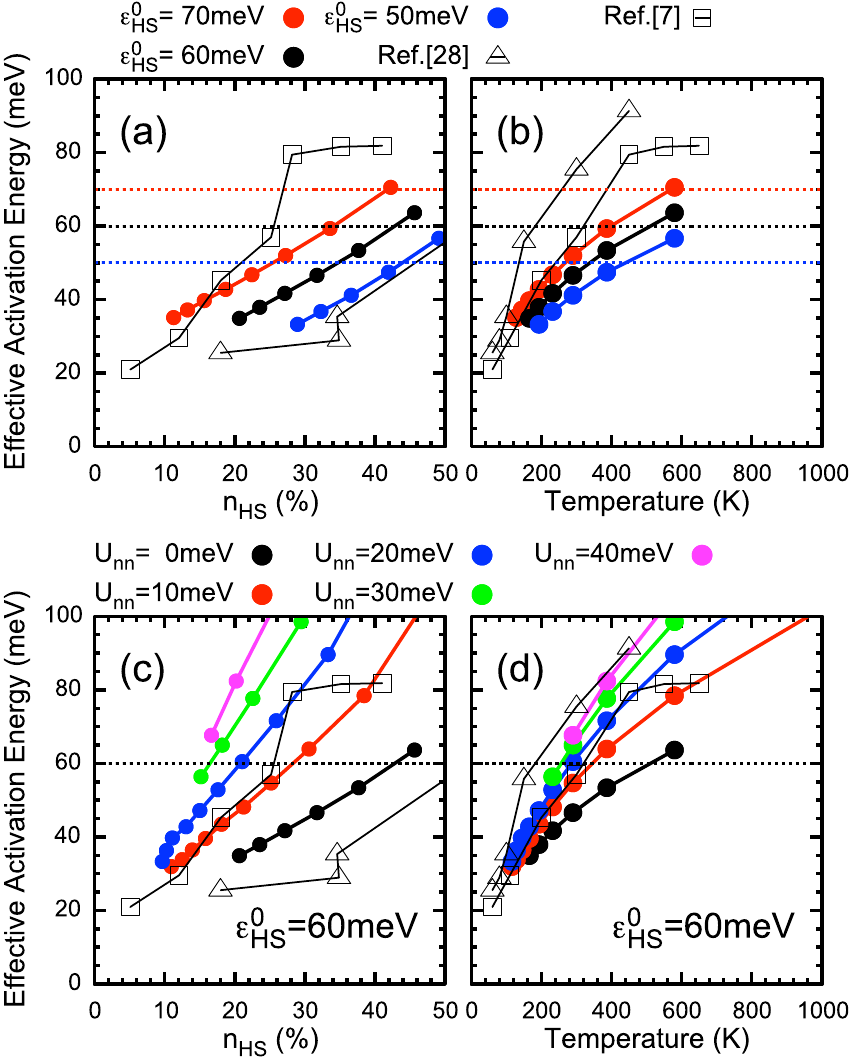}\hspace{0pt}
\caption{The calculated effective activation energies for (a) $n_{\rm HS}$ and (b) temperatures. The different colors correspond to different bare HS energies $\epsilon^0_{\rm HS}$=~70~meV ($V=610$~meV), 60~meV ($V=620$~meV) and ~50~meV ($V=630$~meV), additionally indicated by horizontal dashed lines. The experimental estimates~\cite{Haverkort2006prl,Tomiyasu2017prl} are provided.
(c,d) $\epsilon_{\rm HS}(T)$ at different intersite interactions $U_{\rm nn}$ for $\epsilon^0_{\rm HS}$=60~meV.}
%\caption{
%The effective activation energies according to Eq.~\ref{eq:fd} for the HS populations (a) and temperatures (b) of the HB-DMFT solutions.
% The different colors correspond to different bare HS energies $\epsilon^0_{\rm HS}$=~70~meV ($V=610$~meV), 60~meV ($V=620$~meV) and ~50~meV ($V=630$~meV). The horizontal dashed lines represent these bare energies. The experimental estimates~\cite{Haverkort2006prl,Tomiyasu2017prl} are provided.
%}
\label{fig4}
\end{figure}

As discussed in Refs.~\onlinecite{Sotnikov2016sr,Wang2018prb}, the HS excitation is expected to be almost immobile and thus well described with a $k$-independent excitation energy,
$\epsilon^0_{\rm{HS}}$ in case of an isolated Co atom.
This changes on the lattice.
To quantify this effect, we define a $T$-dependent effective activation energy $\epsilon_{\rm{HS}}(T)$.
It is the excitation energy in a hypothetical isolated atom with the LS ground state and
three excited HS states
(HS$_x$, HS$_y$, HS$_z$),
which leads to the same HS population $n_{\rm{HS}}(T)$ as the lattice model at a given temperature $T$,
%%%%%%%%%%%%%%%%%%%%%%%%%%%%%%%
\begin{equation}
\label{eq:fd}
    \epsilon_{\rm{HS}}(T)=T \operatorname{ln}\frac{3-3n_{\rm{HS}}(T)}{n_{\rm{HS}}(T)}.
\end{equation}
%%%%%%%%%%%%%%%%%%%%%%%%%%%%%%%
%
In Fig.~\ref{fig4} we show $\epsilon_{\rm{HS}}(T)$ as well as $\epsilon_{\rm{HS}}(n_{\rm{HS}})$
dependencies for interaction strength $V$ and compare them to available experimental data~\cite{Haverkort2006prl,Tomiyasu2017prl}.
We find that the low-$T$ value of $\epsilon_{\rm{HS}}$
(an isolated HS excitation on a lattice)
is substantially smaller than $\epsilon^0_{\rm{HS}}$
(HS excitation in an isolated atom).
The explanation of this observation is straightforward.
While the HS excitation is a stable bound state, it is not localized on a single atom. Quantum fluctuations on the adjacent nearest-neighboring bonds of the type $|\rm{HS},\rm{LS}\rangle\rightleftarrows|\rm{IS},\rm{IS}\rangle$,
see Fig.~\ref{fig1},
lower its energy.
A simple perturbation theory gives an estimate $\epsilon^0_{\rm{HS}}-\epsilon_{\rm{HS}}\approx 8{t^2}/{V}
\approx 40$~meV, which describes the calculated data quite well.
It should be pointed out that the on-site energies $\epsilon_{\rm{IS}}$ and $\epsilon^0_{\rm{HS}}$
entering (\ref{eq:model}) are not purely atomic energies, but contain renormalization due to virtual electron hopping~\cite{Afonso2018jccm,Wang2018prb}.

%Comparison to experimental data in Fig.~\ref{fig4}ab leads to two observations. First, the HS concentrations obtained in present study are in a realistic range. Second,
%the effective energy $\epsilon_{\rm{HS}(T)}$ extracted from experiments grows faster with temperature than the theoretical ones. To understand this discrepancy we need to understand why 
%$\epsilon_{\rm{HS}(T)}$ grows. It is understood that this increase 

Similar to the experimental observations, we find that $\epsilon_{\rm{HS}}$ increases with temperature.
The calculated $d\epsilon_{\rm{HS}}/dT$ has a realistic order of magnitude, but is smaller than the experimental observations~\cite{Haverkort2006prl,Tomiyasu2017prl}.
This is not unexpected. The increase of $\epsilon_{\rm{HS}}$ with $n_{\rm{HS}}$ is generally accepted to be caused by nearest-neighbor HS-HS repulsion, which originates from blocking of virtual electron hopping~\cite{Kunes2011prl,Zhang2012,Krapek12,Karolak15} and possibly electron-lattice coupling~\cite{Bari72}.
The strong coupling model of Ref.~\onlinecite{Wang2018prb} captures the former effect. Within the bi-exciton picture
of the HS state, the HS-HS repulsion is inherited from the nearest-neighbor repulsion of the underlying IS excitons.
Similar to their hopping amplitudes (\ref{eq:model}), the IS nearest-neighbor repulsion is strongly anisotropic, inheriting
the anisotropy of the constituting electronic orbitals.

A simple, although approximate, way to include the inter-site repulsion in our model is a static mean-field treatment. 
Following this approach, we consider only the dominant nearest-neighbor repulsion between the excitons of the same
flavor within the their hopping plane
$H_{nn}=U_{nn}\sum_{\bold{i}}\sum_{\alpha,\beta}
(1-\delta_{\alpha \beta} )\hat{n}^{\phantom\dag}_{\bold{i}+\bold{e}_{\beta},\alpha} \hat{n}^{\phantom\dag}_{\bold{i},\alpha}$.
%A approximate way to include the HS-HS repulsion in the DMFT treatment without additional computational cost is a static mean-field approximation of the nearest-neighbor repulsion.
The results in Figs.~\ref{fig4}cd demonstrate that 
inclusion of the inter-site repulsion indeed improves the agreement with experimental estimates. 
The additional term does not change the character of the spectra in Fig.~\ref{fig3}, as it simply introduces a $T$-dependent and self-consistently determined $\epsilon_{\rm{IS}}$ in (\ref{eq:model}).
We point out that the static mean-field approximation
is rather crude as overestimates the effect of the repulsion
especially at low excitation densities, where it ignores the freedom of the excitons to efficiently avoid each other. This is reflected by rather small value of $U_{nn}$ needed, while the strong-coupling estimates are in 100-200~meV range (depending on the spin configuration).

%%%%%%%%%%%%%%%%%%%%%%%%%%%%%%%%%%%%%%%%%%%%%%%%%%%%%%%%%%
%%%%%%%%%%%%%%%%%%%%%%%%%%%%%%%%%%%%%%%%%%%%%%%%%%%%%%%%%%

%Finally, we would like to comment on recent inelastic neutron measurements~\cite{Tomiyasu2018arXiv} reporting
%unusual
%delocalized magnetic form factor
%in LaCoO$_3$ at intermediate temperatures. The present picture of HS excitation as 
%bi-exciton dressed with $|\rm{HS},\rm{LS}\rangle\rightleftarrows|\rm{IS},\rm{IS}\rangle$ on nearest-neighboring bonds
%fits very well with this observation.

In summary, by measuring Co $L_3$-RIXS of LaCoO$_3$ with state of the art energy resolution, we have demonstrated a sizeable mobility of $^3T_{1g}$ IS excitations. 
Melting of the corresponding dispersion into a narrow band of damped excitations, observed between 70 and 150~K, points to a strongly interacting nature of the IS excitations.
The experimental data are well described by 
%modelling the low-energy physics of LaCoO$_3$ with 
a gas of mobile IS excitons and immobile HS bi-excitons, in contrast to the paradigm of atom-bound (immobile) IS or HS excitations.
Our numerical results show that even at low temperatures, where the IS concentration is negligible, virtual fluctuations $|\rm{HS},\rm{LS}\rangle\rightleftarrows|\rm{IS},\rm{IS}\rangle$ on the nearest-neighbor bonds
%enabled by high IS mobility have 
play an important role.
Dressing of HS excitations with a cloud of IS excitations on the neighboring atoms, Fig.~\ref{fig1},
explains the recent inelastic neutron measurements~\cite{Tomiyasu2018arXiv},
which reported delocalized magnetic form factors in LaCoO$_3$ and short range ferromagnetic correlation at intermediate temperatures. Low theoretical IS concentrations obtained 
at elevated temperatures suggest that the equilibrium state of LaCoO$_3$ can be 
described entirely in terms of dressed HS excitations. The IS excitations cannot
be neglected as they mediate the inter-site interactions between these HS excitations.
This conclusion relies on the high mobility of IS excitations and 
is not sensitive to the exact value of IS excitation energy as long as the deviation from the present estimate is smaller than the difference $\epsilon_{\rm{IS}}-\epsilon_{\rm{HS}}$.
The classic cobaltite question whether the low-energy physics is determined by HS or IS excitations is therefore ill posed since both have to be taken into account.

\begin{acknowledgements}
The authors acknowledge A. Kauch, J. Fern\'andez Afonso, D. J. Huang and J. Okamoto for valuable discussions.
The authors thank M. van der Linden and F. Frati for their support in RIXS experiment.
A.H., A.S., and J.K. work are supported by the European Research Council (ERC) under the European Union's Horizon 2020 research and innovation programme (grant agreement No.~646807-EXMAG).
The experiments were supported by ERC advanced grant (grant agreement No.~340279-XRAYonACTIVE). K.T. was financially supported by the MEXT and JSPS KAKENHI (JP17H06137, JP15H03692, JP18K03503).
The calculations were performed on the Vienna Scientific Cluster (VSC).
\end{acknowledgements}

%\textcolor{red}{The appendix was moved to SM, please jump to https://www.overleaf.com/9561149652pqdttvwjycgd.}

\bibliography{main}

\begin{thebibliography}{51}%
\makeatletter
\providecommand \@ifxundefined [1]{%
 \@ifx{#1\undefined}
}%
\providecommand \@ifnum [1]{%
 \ifnum #1\expandafter \@firstoftwo
 \else \expandafter \@secondoftwo
 \fi
}%
\providecommand \@ifx [1]{%
 \ifx #1\expandafter \@firstoftwo
 \else \expandafter \@secondoftwo
 \fi
}%
\providecommand \natexlab [1]{#1}%
\providecommand \enquote  [1]{``#1''}%
\providecommand \bibnamefont  [1]{#1}%
\providecommand \bibfnamefont [1]{#1}%
\providecommand \citenamefont [1]{#1}%
\providecommand \href@noop [0]{\@secondoftwo}%
\providecommand \href [0]{\begingroup \@sanitize@url \@href}%
\providecommand \@href[1]{\@@startlink{#1}\@@href}%
\providecommand \@@href[1]{\endgroup#1\@@endlink}%
\providecommand \@sanitize@url [0]{\catcode `\\12\catcode `\$12\catcode
  `\&12\catcode `\#12\catcode `\^12\catcode `\_12\catcode `\%12\relax}%
\providecommand \@@startlink[1]{}%
\providecommand \@@endlink[0]{}%
\providecommand \url  [0]{\begingroup\@sanitize@url \@url }%
\providecommand \@url [1]{\endgroup\@href {#1}{\urlprefix }}%
\providecommand \urlprefix  [0]{URL }%
\providecommand \Eprint [0]{\href }%
\providecommand \doibase [0]{http://dx.doi.org/}%
\providecommand \selectlanguage [0]{\@gobble}%
\providecommand \bibinfo  [0]{\@secondoftwo}%
\providecommand \bibfield  [0]{\@secondoftwo}%
\providecommand \translation [1]{[#1]}%
\providecommand \BibitemOpen [0]{}%
\providecommand \bibitemStop [0]{}%
\providecommand \bibitemNoStop [0]{.\EOS\space}%
\providecommand \EOS [0]{\spacefactor3000\relax}%
\providecommand \BibitemShut  [1]{\csname bibitem#1\endcsname}%
\let\auto@bib@innerbib\@empty
%</preamble>
\bibitem [{\citenamefont {Goodenough}(1958)}]{Goodenough1958jpcs}%
  \BibitemOpen
  \bibfield  {author} {\bibinfo {author} {\bibfnamefont {J.~B.}\ \bibnamefont
  {Goodenough}},\ }\href {\doibase
  https://doi.org/10.1016/0022-3697(58)90107-0} {\bibfield  {journal} {\bibinfo
   {journal} {J. Phys. Chem. Solid}\ }\textbf {\bibinfo {volume} {6}},\
  \bibinfo {pages} {287} (\bibinfo {year} {1958})}\BibitemShut {NoStop}%
\bibitem [{\citenamefont {Heikes}\ \emph {et~al.}(1964)\citenamefont {Heikes},
  \citenamefont {Miller},\ and\ \citenamefont {Mazelsky}}]{Heikes64}%
  \BibitemOpen
  \bibfield  {author} {\bibinfo {author} {\bibfnamefont {R.}~\bibnamefont
  {Heikes}}, \bibinfo {author} {\bibfnamefont {R.}~\bibnamefont {Miller}}, \
  and\ \bibinfo {author} {\bibfnamefont {R.}~\bibnamefont {Mazelsky}},\ }\href
  {\doibase https://doi.org/10.1016/0031-8914(64)90182-X} {\bibfield  {journal}
  {\bibinfo  {journal} {Physica}\ }\textbf {\bibinfo {volume} {30}},\ \bibinfo
  {pages} {1600 } (\bibinfo {year} {1964})}\BibitemShut {NoStop}%
\bibitem [{\citenamefont {Raccah}\ and\ \citenamefont
  {Goodenough}(1967)}]{Raccah1967pr}%
  \BibitemOpen
  \bibfield  {author} {\bibinfo {author} {\bibfnamefont {P.~M.}\ \bibnamefont
  {Raccah}}\ and\ \bibinfo {author} {\bibfnamefont {J.~B.}\ \bibnamefont
  {Goodenough}},\ }\href {\doibase 10.1103/PhysRev.155.932} {\bibfield
  {journal} {\bibinfo  {journal} {Phys. Rev.}\ }\textbf {\bibinfo {volume}
  {155}},\ \bibinfo {pages} {932} (\bibinfo {year} {1967})}\BibitemShut
  {NoStop}%
\bibitem [{\citenamefont {Abbate}\ \emph {et~al.}(1993)\citenamefont {Abbate},
  \citenamefont {Fuggle}, \citenamefont {Fujimori}, \citenamefont {Tjeng},
  \citenamefont {Chen}, \citenamefont {Potze}, \citenamefont {Sawatzky},
  \citenamefont {Eisaki},\ and\ \citenamefont {Uchida}}]{Abbate1993prb}%
  \BibitemOpen
  \bibfield  {author} {\bibinfo {author} {\bibfnamefont {M.}~\bibnamefont
  {Abbate}}, \bibinfo {author} {\bibfnamefont {J.~C.}\ \bibnamefont {Fuggle}},
  \bibinfo {author} {\bibfnamefont {A.}~\bibnamefont {Fujimori}}, \bibinfo
  {author} {\bibfnamefont {L.~H.}\ \bibnamefont {Tjeng}}, \bibinfo {author}
  {\bibfnamefont {C.~T.}\ \bibnamefont {Chen}}, \bibinfo {author}
  {\bibfnamefont {R.}~\bibnamefont {Potze}}, \bibinfo {author} {\bibfnamefont
  {G.~A.}\ \bibnamefont {Sawatzky}}, \bibinfo {author} {\bibfnamefont
  {H.}~\bibnamefont {Eisaki}}, \ and\ \bibinfo {author} {\bibfnamefont
  {S.}~\bibnamefont {Uchida}},\ }\href {\doibase 10.1103/PhysRevB.47.16124}
  {\bibfield  {journal} {\bibinfo  {journal} {Phys. Rev. B}\ }\textbf {\bibinfo
  {volume} {47}},\ \bibinfo {pages} {16124} (\bibinfo {year}
  {1993})}\BibitemShut {NoStop}%
\bibitem [{\citenamefont {Asai}\ \emph {et~al.}(1998)\citenamefont {Asai},
  \citenamefont {Yoneda}, \citenamefont {Yokokura}, \citenamefont {Tranquada},
  \citenamefont {Shirane},\ and\ \citenamefont {Kohn}}]{Asai1998jpsj}%
  \BibitemOpen
  \bibfield  {author} {\bibinfo {author} {\bibfnamefont {K.}~\bibnamefont
  {Asai}}, \bibinfo {author} {\bibfnamefont {A.}~\bibnamefont {Yoneda}},
  \bibinfo {author} {\bibfnamefont {O.}~\bibnamefont {Yokokura}}, \bibinfo
  {author} {\bibfnamefont {J.~M.}\ \bibnamefont {Tranquada}}, \bibinfo {author}
  {\bibfnamefont {G.}~\bibnamefont {Shirane}}, \ and\ \bibinfo {author}
  {\bibfnamefont {K.}~\bibnamefont {Kohn}},\ }\href {\doibase
  10.1143/JPSJ.67.290} {\bibfield  {journal} {\bibinfo  {journal} {J. Phys.
  Soc. Jpn.}\ }\textbf {\bibinfo {volume} {67}},\ \bibinfo {pages} {290}
  (\bibinfo {year} {1998})}\BibitemShut {NoStop}%
\bibitem [{\citenamefont {St\o{}len}\ \emph {et~al.}(1997)\citenamefont
  {St\o{}len}, \citenamefont {Gr\o{}nvold}, \citenamefont {Brinks},
  \citenamefont {Atake},\ and\ \citenamefont {Mori}}]{Stolen1997prb}%
  \BibitemOpen
  \bibfield  {author} {\bibinfo {author} {\bibfnamefont {S.}~\bibnamefont
  {St\o{}len}}, \bibinfo {author} {\bibfnamefont {F.}~\bibnamefont
  {Gr\o{}nvold}}, \bibinfo {author} {\bibfnamefont {H.}~\bibnamefont {Brinks}},
  \bibinfo {author} {\bibfnamefont {T.}~\bibnamefont {Atake}}, \ and\ \bibinfo
  {author} {\bibfnamefont {H.}~\bibnamefont {Mori}},\ }\href {\doibase
  10.1103/PhysRevB.55.14103} {\bibfield  {journal} {\bibinfo  {journal} {Phys.
  Rev. B}\ }\textbf {\bibinfo {volume} {55}},\ \bibinfo {pages} {14103}
  (\bibinfo {year} {1997})}\BibitemShut {NoStop}%
\bibitem [{\citenamefont {Haverkort}\ \emph {et~al.}(2006)\citenamefont
  {Haverkort}, \citenamefont {Hu}, \citenamefont {Cezar}, \citenamefont
  {Burnus}, \citenamefont {Hartmann}, \citenamefont {Reuther}, \citenamefont
  {Zobel}, \citenamefont {Lorenz}, \citenamefont {Tanaka}, \citenamefont
  {Brookes}, \citenamefont {Hsieh}, \citenamefont {Lin}, \citenamefont {Chen},\
  and\ \citenamefont {Tjeng}}]{Haverkort2006prl}%
  \BibitemOpen
  \bibfield  {author} {\bibinfo {author} {\bibfnamefont {M.~W.}\ \bibnamefont
  {Haverkort}}, \bibinfo {author} {\bibfnamefont {Z.}~\bibnamefont {Hu}},
  \bibinfo {author} {\bibfnamefont {J.~C.}\ \bibnamefont {Cezar}}, \bibinfo
  {author} {\bibfnamefont {T.}~\bibnamefont {Burnus}}, \bibinfo {author}
  {\bibfnamefont {H.}~\bibnamefont {Hartmann}}, \bibinfo {author}
  {\bibfnamefont {M.}~\bibnamefont {Reuther}}, \bibinfo {author} {\bibfnamefont
  {C.}~\bibnamefont {Zobel}}, \bibinfo {author} {\bibfnamefont
  {T.}~\bibnamefont {Lorenz}}, \bibinfo {author} {\bibfnamefont
  {A.}~\bibnamefont {Tanaka}}, \bibinfo {author} {\bibfnamefont {N.~B.}\
  \bibnamefont {Brookes}}, \bibinfo {author} {\bibfnamefont {H.~H.}\
  \bibnamefont {Hsieh}}, \bibinfo {author} {\bibfnamefont {H.-J.}\ \bibnamefont
  {Lin}}, \bibinfo {author} {\bibfnamefont {C.~T.}\ \bibnamefont {Chen}}, \
  and\ \bibinfo {author} {\bibfnamefont {L.~H.}\ \bibnamefont {Tjeng}},\ }\href
  {\doibase 10.1103/PhysRevLett.97.176405} {\bibfield  {journal} {\bibinfo
  {journal} {Phys. Rev. Lett.}\ }\textbf {\bibinfo {volume} {97}},\ \bibinfo
  {pages} {176405} (\bibinfo {year} {2006})}\BibitemShut {NoStop}%
\bibitem [{\citenamefont {Tanabe}\ and\ \citenamefont
  {Sugano}(1954)}]{Tanabe1954jpcj}%
  \BibitemOpen
  \bibfield  {author} {\bibinfo {author} {\bibfnamefont {Y.}~\bibnamefont
  {Tanabe}}\ and\ \bibinfo {author} {\bibfnamefont {S.}~\bibnamefont
  {Sugano}},\ }\href@noop {} {\bibfield  {journal} {\bibinfo  {journal} {J.
  Phys. Soc. Jpn.}\ }\textbf {\bibinfo {volume} {9}},\ \bibinfo {pages} {766}
  (\bibinfo {year} {1954})}\BibitemShut {NoStop}%
\bibitem [{\citenamefont {de~Groot}\ \emph {et~al.}(1990)\citenamefont
  {de~Groot}, \citenamefont {Fuggle}, \citenamefont {Thole},\ and\
  \citenamefont {Sawatzky}}]{deGroot1990prb}%
  \BibitemOpen
  \bibfield  {author} {\bibinfo {author} {\bibfnamefont {F.~M.~F.}\
  \bibnamefont {de~Groot}}, \bibinfo {author} {\bibfnamefont {J.~C.}\
  \bibnamefont {Fuggle}}, \bibinfo {author} {\bibfnamefont {B.~T.}\
  \bibnamefont {Thole}}, \ and\ \bibinfo {author} {\bibfnamefont {G.~A.}\
  \bibnamefont {Sawatzky}},\ }\href {\doibase 10.1103/PhysRevB.42.5459}
  {\bibfield  {journal} {\bibinfo  {journal} {Phys. Rev. B}\ }\textbf {\bibinfo
  {volume} {42}},\ \bibinfo {pages} {5459} (\bibinfo {year}
  {1990})}\BibitemShut {NoStop}%
\bibitem [{\citenamefont {Zobel}\ \emph {et~al.}(2002)\citenamefont {Zobel},
  \citenamefont {Kriener}, \citenamefont {Bruns}, \citenamefont {Baier},
  \citenamefont {Gr\"uninger}, \citenamefont {Lorenz}, \citenamefont
  {Reutler},\ and\ \citenamefont {Revcolevschi}}]{Zobel2002prb}%
  \BibitemOpen
  \bibfield  {author} {\bibinfo {author} {\bibfnamefont {C.}~\bibnamefont
  {Zobel}}, \bibinfo {author} {\bibfnamefont {M.}~\bibnamefont {Kriener}},
  \bibinfo {author} {\bibfnamefont {D.}~\bibnamefont {Bruns}}, \bibinfo
  {author} {\bibfnamefont {J.}~\bibnamefont {Baier}}, \bibinfo {author}
  {\bibfnamefont {M.}~\bibnamefont {Gr\"uninger}}, \bibinfo {author}
  {\bibfnamefont {T.}~\bibnamefont {Lorenz}}, \bibinfo {author} {\bibfnamefont
  {P.}~\bibnamefont {Reutler}}, \ and\ \bibinfo {author} {\bibfnamefont
  {A.}~\bibnamefont {Revcolevschi}},\ }\href {\doibase
  10.1103/PhysRevB.66.020402} {\bibfield  {journal} {\bibinfo  {journal} {Phys.
  Rev. B}\ }\textbf {\bibinfo {volume} {66}},\ \bibinfo {pages} {020402}
  (\bibinfo {year} {2002})}\BibitemShut {NoStop}%
\bibitem [{\citenamefont {Ishikawa}\ \emph {et~al.}(2004)\citenamefont
  {Ishikawa}, \citenamefont {Nohara},\ and\ \citenamefont
  {Sugai}}]{Ishikawa2004prl}%
  \BibitemOpen
  \bibfield  {author} {\bibinfo {author} {\bibfnamefont {A.}~\bibnamefont
  {Ishikawa}}, \bibinfo {author} {\bibfnamefont {J.}~\bibnamefont {Nohara}}, \
  and\ \bibinfo {author} {\bibfnamefont {S.}~\bibnamefont {Sugai}},\ }\href
  {\doibase 10.1103/PhysRevLett.93.136401} {\bibfield  {journal} {\bibinfo
  {journal} {Phys. Rev. Lett.}\ }\textbf {\bibinfo {volume} {93}},\ \bibinfo
  {pages} {136401} (\bibinfo {year} {2004})}\BibitemShut {NoStop}%
\bibitem [{\citenamefont {Korotin}\ \emph {et~al.}(1996)\citenamefont
  {Korotin}, \citenamefont {Ezhov}, \citenamefont {Solovyev}, \citenamefont
  {Anisimov}, \citenamefont {Khomskii},\ and\ \citenamefont
  {Sawatzky}}]{Korotin1996prb}%
  \BibitemOpen
  \bibfield  {author} {\bibinfo {author} {\bibfnamefont {M.~A.}\ \bibnamefont
  {Korotin}}, \bibinfo {author} {\bibfnamefont {S.~Y.}\ \bibnamefont {Ezhov}},
  \bibinfo {author} {\bibfnamefont {I.~V.}\ \bibnamefont {Solovyev}}, \bibinfo
  {author} {\bibfnamefont {V.~I.}\ \bibnamefont {Anisimov}}, \bibinfo {author}
  {\bibfnamefont {D.~I.}\ \bibnamefont {Khomskii}}, \ and\ \bibinfo {author}
  {\bibfnamefont {G.~A.}\ \bibnamefont {Sawatzky}},\ }\href {\doibase
  10.1103/PhysRevB.54.5309} {\bibfield  {journal} {\bibinfo  {journal} {Phys.
  Rev. B}\ }\textbf {\bibinfo {volume} {54}},\ \bibinfo {pages} {5309}
  (\bibinfo {year} {1996})}\BibitemShut {NoStop}%
\bibitem [{\citenamefont {Yan}\ \emph {et~al.}(2004)\citenamefont {Yan},
  \citenamefont {Zhou},\ and\ \citenamefont {Goodenough}}]{Yan2004prb}%
  \BibitemOpen
  \bibfield  {author} {\bibinfo {author} {\bibfnamefont {J.-Q.}\ \bibnamefont
  {Yan}}, \bibinfo {author} {\bibfnamefont {J.~S.}\ \bibnamefont {Zhou}}, \
  and\ \bibinfo {author} {\bibfnamefont {J.~B.}\ \bibnamefont {Goodenough}},\
  }\href {\doibase 10.1103/PhysRevB.69.134409} {\bibfield  {journal} {\bibinfo
  {journal} {Phys. Rev. B.}\ }\textbf {\bibinfo {volume} {69}},\ \bibinfo
  {pages} {134409} (\bibinfo {year} {2004})}\BibitemShut {NoStop}%
\bibitem [{\citenamefont {Ropka}\ and\ \citenamefont
  {Radwanski}(2003)}]{Ropka2003prb}%
  \BibitemOpen
  \bibfield  {author} {\bibinfo {author} {\bibfnamefont {Z.}~\bibnamefont
  {Ropka}}\ and\ \bibinfo {author} {\bibfnamefont {R.~J.}\ \bibnamefont
  {Radwanski}},\ }\href {\doibase 10.1103/PhysRevB.67.172401} {\bibfield
  {journal} {\bibinfo  {journal} {Phys. Rev. B}\ }\textbf {\bibinfo {volume}
  {67}},\ \bibinfo {pages} {172401} (\bibinfo {year} {2003})}\BibitemShut
  {NoStop}%
\bibitem [{\citenamefont {Podlesnyak}\ \emph {et~al.}(2006)\citenamefont
  {Podlesnyak}, \citenamefont {Streule}, \citenamefont {Mesot}, \citenamefont
  {Medarde}, \citenamefont {Pomjakushina}, \citenamefont {Conder},
  \citenamefont {Tanaka}, \citenamefont {Haverkort},\ and\ \citenamefont
  {Khomskii}}]{Podlesnyak2006prl}%
  \BibitemOpen
  \bibfield  {author} {\bibinfo {author} {\bibfnamefont {A.}~\bibnamefont
  {Podlesnyak}}, \bibinfo {author} {\bibfnamefont {S.}~\bibnamefont {Streule}},
  \bibinfo {author} {\bibfnamefont {J.}~\bibnamefont {Mesot}}, \bibinfo
  {author} {\bibfnamefont {M.}~\bibnamefont {Medarde}}, \bibinfo {author}
  {\bibfnamefont {E.}~\bibnamefont {Pomjakushina}}, \bibinfo {author}
  {\bibfnamefont {K.}~\bibnamefont {Conder}}, \bibinfo {author} {\bibfnamefont
  {A.}~\bibnamefont {Tanaka}}, \bibinfo {author} {\bibfnamefont {M.~W.}\
  \bibnamefont {Haverkort}}, \ and\ \bibinfo {author} {\bibfnamefont {D.~I.}\
  \bibnamefont {Khomskii}},\ }\href {\doibase 10.1103/PhysRevLett.97.247208}
  {\bibfield  {journal} {\bibinfo  {journal} {Phys. Rev. Lett.}\ }\textbf
  {\bibinfo {volume} {97}},\ \bibinfo {pages} {247208} (\bibinfo {year}
  {2006})}\BibitemShut {NoStop}%
\bibitem [{\citenamefont {Noguchi}\ \emph {et~al.}(2002)\citenamefont
  {Noguchi}, \citenamefont {Kawamata}, \citenamefont {Okuda}, \citenamefont
  {Nojiri},\ and\ \citenamefont {Motokawa}}]{Noguchi2002prb}%
  \BibitemOpen
  \bibfield  {author} {\bibinfo {author} {\bibfnamefont {S.}~\bibnamefont
  {Noguchi}}, \bibinfo {author} {\bibfnamefont {S.}~\bibnamefont {Kawamata}},
  \bibinfo {author} {\bibfnamefont {K.}~\bibnamefont {Okuda}}, \bibinfo
  {author} {\bibfnamefont {H.}~\bibnamefont {Nojiri}}, \ and\ \bibinfo {author}
  {\bibfnamefont {M.}~\bibnamefont {Motokawa}},\ }\href {\doibase
  10.1103/PhysRevB.66.094404} {\bibfield  {journal} {\bibinfo  {journal} {Phys.
  Rev. B}\ }\textbf {\bibinfo {volume} {66}},\ \bibinfo {pages} {094404}
  (\bibinfo {year} {2002})}\BibitemShut {NoStop}%
\bibitem [{\citenamefont {Maris}\ \emph {et~al.}(2003)\citenamefont {Maris},
  \citenamefont {Ren}, \citenamefont {Volotchaev}, \citenamefont {Zobel},
  \citenamefont {Lorenz},\ and\ \citenamefont {Palstra}}]{Maris2003prb}%
  \BibitemOpen
  \bibfield  {author} {\bibinfo {author} {\bibfnamefont {G.}~\bibnamefont
  {Maris}}, \bibinfo {author} {\bibfnamefont {Y.}~\bibnamefont {Ren}}, \bibinfo
  {author} {\bibfnamefont {V.}~\bibnamefont {Volotchaev}}, \bibinfo {author}
  {\bibfnamefont {C.}~\bibnamefont {Zobel}}, \bibinfo {author} {\bibfnamefont
  {T.}~\bibnamefont {Lorenz}}, \ and\ \bibinfo {author} {\bibfnamefont
  {T.~T.~M.}\ \bibnamefont {Palstra}},\ }\href {\doibase
  10.1103/PhysRevB.67.224423} {\bibfield  {journal} {\bibinfo  {journal} {Phys.
  Rev. B}\ }\textbf {\bibinfo {volume} {67}},\ \bibinfo {pages} {224423}
  (\bibinfo {year} {2003})}\BibitemShut {NoStop}%
\bibitem [{\citenamefont {Vogt}\ \emph {et~al.}(2003)\citenamefont {Vogt},
  \citenamefont {Hriljac}, \citenamefont {Hyatt},\ and\ \citenamefont
  {Woodward}}]{Vogt2003prb}%
  \BibitemOpen
  \bibfield  {author} {\bibinfo {author} {\bibfnamefont {T.}~\bibnamefont
  {Vogt}}, \bibinfo {author} {\bibfnamefont {J.~A.}\ \bibnamefont {Hriljac}},
  \bibinfo {author} {\bibfnamefont {N.~C.}\ \bibnamefont {Hyatt}}, \ and\
  \bibinfo {author} {\bibfnamefont {P.}~\bibnamefont {Woodward}},\ }\href
  {\doibase 10.1103/PhysRevB.67.140401} {\bibfield  {journal} {\bibinfo
  {journal} {Phys. Rev. B}\ }\textbf {\bibinfo {volume} {67}},\ \bibinfo
  {pages} {140401} (\bibinfo {year} {2003})}\BibitemShut {NoStop}%
\bibitem [{\citenamefont {Saitoh}\ \emph {et~al.}(1997)\citenamefont {Saitoh},
  \citenamefont {Mizokawa}, \citenamefont {Fujimori}, \citenamefont {Abbate},
  \citenamefont {Takeda},\ and\ \citenamefont {Takano}}]{Saitoh1997prb}%
  \BibitemOpen
  \bibfield  {author} {\bibinfo {author} {\bibfnamefont {T.}~\bibnamefont
  {Saitoh}}, \bibinfo {author} {\bibfnamefont {T.}~\bibnamefont {Mizokawa}},
  \bibinfo {author} {\bibfnamefont {A.}~\bibnamefont {Fujimori}}, \bibinfo
  {author} {\bibfnamefont {M.}~\bibnamefont {Abbate}}, \bibinfo {author}
  {\bibfnamefont {Y.}~\bibnamefont {Takeda}}, \ and\ \bibinfo {author}
  {\bibfnamefont {M.}~\bibnamefont {Takano}},\ }\href {\doibase
  10.1103/PhysRevB.55.4257} {\bibfield  {journal} {\bibinfo  {journal} {Phys.
  Rev. B}\ }\textbf {\bibinfo {volume} {55}},\ \bibinfo {pages} {4257}
  (\bibinfo {year} {1997})}\BibitemShut {NoStop}%
\bibitem [{\citenamefont {Bari}\ and\ \citenamefont
  {Sivardi\`ere}(1972)}]{Bari72}%
  \BibitemOpen
  \bibfield  {author} {\bibinfo {author} {\bibfnamefont {R.~A.}\ \bibnamefont
  {Bari}}\ and\ \bibinfo {author} {\bibfnamefont {J.}~\bibnamefont
  {Sivardi\`ere}},\ }\href {\doibase 10.1103/PhysRevB.5.4466} {\bibfield
  {journal} {\bibinfo  {journal} {Phys. Rev. B}\ }\textbf {\bibinfo {volume}
  {5}},\ \bibinfo {pages} {4466} (\bibinfo {year} {1972})}\BibitemShut
  {NoStop}%
\bibitem [{\citenamefont {Kn\'{\i}\ifmmode\check{z}\else\v{z}\fi{}ek}\ \emph
  {et~al.}(2009)\citenamefont {Kn\'{\i}\ifmmode\check{z}\else\v{z}\fi{}ek},
  \citenamefont {Jir\'ak}, \citenamefont {Hejtm\'anek}, \citenamefont
  {Nov\'ak},\ and\ \citenamefont {Ku}}]{Knizek09}%
  \BibitemOpen
  \bibfield  {author} {\bibinfo {author} {\bibfnamefont {K.}~\bibnamefont
  {Kn\'{\i}\ifmmode\check{z}\else\v{z}\fi{}ek}}, \bibinfo {author}
  {\bibfnamefont {Z.}~\bibnamefont {Jir\'ak}}, \bibinfo {author} {\bibfnamefont
  {J.}~\bibnamefont {Hejtm\'anek}}, \bibinfo {author} {\bibfnamefont
  {P.}~\bibnamefont {Nov\'ak}}, \ and\ \bibinfo {author} {\bibfnamefont
  {W.}~\bibnamefont {Ku}},\ }\href {\doibase 10.1103/PhysRevB.79.014430}
  {\bibfield  {journal} {\bibinfo  {journal} {Phys. Rev. B}\ }\textbf {\bibinfo
  {volume} {79}},\ \bibinfo {pages} {014430} (\bibinfo {year}
  {2009})}\BibitemShut {NoStop}%
\bibitem [{\citenamefont {Kune\v{s}}\ and\ \citenamefont
  {K\v{r}\'apek}(2011)}]{Kunes2011prl}%
  \BibitemOpen
  \bibfield  {author} {\bibinfo {author} {\bibfnamefont {J.}~\bibnamefont
  {Kune\v{s}}}\ and\ \bibinfo {author} {\bibfnamefont {V.}~\bibnamefont
  {K\v{r}\'apek}},\ }\href {\doibase 10.1103/PhysRevLett.106.256401} {\bibfield
   {journal} {\bibinfo  {journal} {Phys. Rev. Lett.}\ }\textbf {\bibinfo
  {volume} {106}},\ \bibinfo {pages} {256401} (\bibinfo {year}
  {2011})}\BibitemShut {NoStop}%
\bibitem [{\citenamefont {Karolak}\ \emph {et~al.}(2015)\citenamefont
  {Karolak}, \citenamefont {Izquierdo}, \citenamefont {Molodtsov},\ and\
  \citenamefont {Lichtenstein}}]{Karolak15}%
  \BibitemOpen
  \bibfield  {author} {\bibinfo {author} {\bibfnamefont {M.}~\bibnamefont
  {Karolak}}, \bibinfo {author} {\bibfnamefont {M.}~\bibnamefont {Izquierdo}},
  \bibinfo {author} {\bibfnamefont {S.~L.}\ \bibnamefont {Molodtsov}}, \ and\
  \bibinfo {author} {\bibfnamefont {A.~I.}\ \bibnamefont {Lichtenstein}},\
  }\href {\doibase 10.1103/PhysRevLett.115.046401} {\bibfield  {journal}
  {\bibinfo  {journal} {Phys. Rev. Lett.}\ }\textbf {\bibinfo {volume} {115}},\
  \bibinfo {pages} {046401} (\bibinfo {year} {2015})}\BibitemShut {NoStop}%
\bibitem [{\citenamefont {Zhang}\ \emph
  {et~al.}(2012{\natexlab{a}})\citenamefont {Zhang}, \citenamefont {Gorelov},
  \citenamefont {Koch},\ and\ \citenamefont {Pavarini}}]{Zhang12}%
  \BibitemOpen
  \bibfield  {author} {\bibinfo {author} {\bibfnamefont {G.}~\bibnamefont
  {Zhang}}, \bibinfo {author} {\bibfnamefont {E.}~\bibnamefont {Gorelov}},
  \bibinfo {author} {\bibfnamefont {E.}~\bibnamefont {Koch}}, \ and\ \bibinfo
  {author} {\bibfnamefont {E.}~\bibnamefont {Pavarini}},\ }\href {\doibase
  10.1103/PhysRevB.86.184413} {\bibfield  {journal} {\bibinfo  {journal} {Phys.
  Rev. B}\ }\textbf {\bibinfo {volume} {86}},\ \bibinfo {pages} {184413}
  (\bibinfo {year} {2012}{\natexlab{a}})}\BibitemShut {NoStop}%
\bibitem [{\citenamefont {Radaelli}\ and\ \citenamefont
  {Cheong}(2002)}]{Radaelli02}%
  \BibitemOpen
  \bibfield  {author} {\bibinfo {author} {\bibfnamefont {P.~G.}\ \bibnamefont
  {Radaelli}}\ and\ \bibinfo {author} {\bibfnamefont {S.-W.}\ \bibnamefont
  {Cheong}},\ }\href {\doibase 10.1103/PhysRevB.66.094408} {\bibfield
  {journal} {\bibinfo  {journal} {Phys. Rev. B}\ }\textbf {\bibinfo {volume}
  {66}},\ \bibinfo {pages} {094408} (\bibinfo {year} {2002})}\BibitemShut
  {NoStop}%
\bibitem [{\citenamefont {Doi}\ \emph {et~al.}(2014)\citenamefont {Doi},
  \citenamefont {Fujioka}, \citenamefont {Fukuda}, \citenamefont {Tsutsui},
  \citenamefont {Okuyama}, \citenamefont {Taguchi}, \citenamefont {Arima},
  \citenamefont {Baron},\ and\ \citenamefont {Tokura}}]{Doi2014prb}%
  \BibitemOpen
  \bibfield  {author} {\bibinfo {author} {\bibfnamefont {A.}~\bibnamefont
  {Doi}}, \bibinfo {author} {\bibfnamefont {J.}~\bibnamefont {Fujioka}},
  \bibinfo {author} {\bibfnamefont {T.}~\bibnamefont {Fukuda}}, \bibinfo
  {author} {\bibfnamefont {S.}~\bibnamefont {Tsutsui}}, \bibinfo {author}
  {\bibfnamefont {D.}~\bibnamefont {Okuyama}}, \bibinfo {author} {\bibfnamefont
  {Y.}~\bibnamefont {Taguchi}}, \bibinfo {author} {\bibfnamefont
  {T.}~\bibnamefont {Arima}}, \bibinfo {author} {\bibfnamefont {A.~Q.~R.}\
  \bibnamefont {Baron}}, \ and\ \bibinfo {author} {\bibfnamefont
  {Y.}~\bibnamefont {Tokura}},\ }\href {\doibase 10.1103/PhysRevB.90.081109}
  {\bibfield  {journal} {\bibinfo  {journal} {Phys. Rev. B}\ }\textbf {\bibinfo
  {volume} {90}},\ \bibinfo {pages} {081109} (\bibinfo {year}
  {2014})}\BibitemShut {NoStop}%
\bibitem [{\citenamefont {Sotnikov}\ and\ \citenamefont
  {Kune\v{s}}(2016)}]{Sotnikov2016sr}%
  \BibitemOpen
  \bibfield  {author} {\bibinfo {author} {\bibfnamefont {A.}~\bibnamefont
  {Sotnikov}}\ and\ \bibinfo {author} {\bibfnamefont {J.}~\bibnamefont
  {Kune\v{s}}},\ }\href {\doibase 10.1038/srep30510} {\bibfield  {journal}
  {\bibinfo  {journal} {Sci. Rep.}\ }\textbf {\bibinfo {volume} {6}},\ \bibinfo
  {pages} {30510} (\bibinfo {year} {2016})}\BibitemShut {NoStop}%
\bibitem [{\citenamefont {Tomiyasu}\ \emph {et~al.}(2017)\citenamefont
  {Tomiyasu}, \citenamefont {Okamoto}, \citenamefont {Huang}, \citenamefont
  {Chen}, \citenamefont {Sinaga}, \citenamefont {Wu}, \citenamefont {Chu},
  \citenamefont {Singh}, \citenamefont {Wang}, \citenamefont {de~Groot},
  \citenamefont {Chainani}, \citenamefont {Ishihara}, \citenamefont {Chen},\
  and\ \citenamefont {Huang}}]{Tomiyasu2017prl}%
  \BibitemOpen
  \bibfield  {author} {\bibinfo {author} {\bibfnamefont {K.}~\bibnamefont
  {Tomiyasu}}, \bibinfo {author} {\bibfnamefont {J.}~\bibnamefont {Okamoto}},
  \bibinfo {author} {\bibfnamefont {H.~Y.}\ \bibnamefont {Huang}}, \bibinfo
  {author} {\bibfnamefont {Z.~Y.}\ \bibnamefont {Chen}}, \bibinfo {author}
  {\bibfnamefont {E.~P.}\ \bibnamefont {Sinaga}}, \bibinfo {author}
  {\bibfnamefont {W.~B.}\ \bibnamefont {Wu}}, \bibinfo {author} {\bibfnamefont
  {Y.~Y.}\ \bibnamefont {Chu}}, \bibinfo {author} {\bibfnamefont
  {A.}~\bibnamefont {Singh}}, \bibinfo {author} {\bibfnamefont {R.-P.}\
  \bibnamefont {Wang}}, \bibinfo {author} {\bibfnamefont {F.~M.~F.}\
  \bibnamefont {de~Groot}}, \bibinfo {author} {\bibfnamefont {A.}~\bibnamefont
  {Chainani}}, \bibinfo {author} {\bibfnamefont {S.}~\bibnamefont {Ishihara}},
  \bibinfo {author} {\bibfnamefont {C.~T.}\ \bibnamefont {Chen}}, \ and\
  \bibinfo {author} {\bibfnamefont {D.~J.}\ \bibnamefont {Huang}},\ }\href
  {\doibase 10.1103/PhysRevLett.119.196402} {\bibfield  {journal} {\bibinfo
  {journal} {Phys. Rev. Lett.}\ }\textbf {\bibinfo {volume} {119}},\ \bibinfo
  {pages} {196402} (\bibinfo {year} {2017})}\BibitemShut {NoStop}%
\bibitem [{\citenamefont {Wang}\ \emph {et~al.}(2018)\citenamefont {Wang},
  \citenamefont {Hariki}, \citenamefont {Sotnikov}, \citenamefont {Frati},
  \citenamefont {Okamoto}, \citenamefont {Huang}, \citenamefont {Singh},
  \citenamefont {Huang}, \citenamefont {Tomiyasu}, \citenamefont {Du},\ and\
  \citenamefont {\textit{et. al.}}}]{Wang2018prb}%
  \BibitemOpen
  \bibfield  {author} {\bibinfo {author} {\bibfnamefont {R.-P.}\ \bibnamefont
  {Wang}}, \bibinfo {author} {\bibfnamefont {A.}~\bibnamefont {Hariki}},
  \bibinfo {author} {\bibfnamefont {A.}~\bibnamefont {Sotnikov}}, \bibinfo
  {author} {\bibfnamefont {F.}~\bibnamefont {Frati}}, \bibinfo {author}
  {\bibfnamefont {J.}~\bibnamefont {Okamoto}}, \bibinfo {author} {\bibfnamefont
  {H.}~\bibnamefont {Huang}}, \bibinfo {author} {\bibfnamefont
  {A.}~\bibnamefont {Singh}}, \bibinfo {author} {\bibfnamefont
  {D.}~\bibnamefont {Huang}}, \bibinfo {author} {\bibfnamefont
  {K.}~\bibnamefont {Tomiyasu}}, \bibinfo {author} {\bibfnamefont
  {C.}~\bibnamefont {Du}}, \ and\ \bibinfo {author} {\bibnamefont {\textit{et.
  al.}}},\ }\href {\doibase 10.1103/PhysRevB.98.035149} {\bibfield  {journal}
  {\bibinfo  {journal} {Phys. Rev. B}\ }\textbf {\bibinfo {volume} {98}},\
  \bibinfo {pages} {035149} (\bibinfo {year} {2018})}\BibitemShut {NoStop}%
\bibitem [{sm()}]{sm}%
  \BibitemOpen
  \href@noop {} {\ }\bibinfo {note} {See the Supplementary Material for
  computational details and fitting analysis of the experimental data, which
  includes
  Refs.~\cite{wien2wannier,wannier90,Ikeda2016prb,Afonso19,jarrell96,Gull11}.}\BibitemShut
  {Stop}%
\bibitem [{\citenamefont {Brookes}\ \emph {et~al.}(2018)\citenamefont
  {Brookes}, \citenamefont {Yakhou-Harris}, \citenamefont {Kummer},
  \citenamefont {Fondacaro}, \citenamefont {Cezar}, \citenamefont {Betto},
  \citenamefont {Velez-Fort}, \citenamefont {Amorese}, \citenamefont
  {Ghiringhelli}, \citenamefont {Braicovich}, \citenamefont {Barrett},
  \citenamefont {Berruyer}, \citenamefont {Cianciosi}, \citenamefont {Eybert},
  \citenamefont {Marion}, \citenamefont {van~der Linden},\ and\ \citenamefont
  {Zhang}}]{Brookes18}%
  \BibitemOpen
  \bibfield  {author} {\bibinfo {author} {\bibfnamefont {N.~B.}\ \bibnamefont
  {Brookes}}, \bibinfo {author} {\bibfnamefont {F.}~\bibnamefont
  {Yakhou-Harris}}, \bibinfo {author} {\bibfnamefont {K.}~\bibnamefont
  {Kummer}}, \bibinfo {author} {\bibfnamefont {A.}~\bibnamefont {Fondacaro}},
  \bibinfo {author} {\bibfnamefont {J.~C.}\ \bibnamefont {Cezar}}, \bibinfo
  {author} {\bibfnamefont {D.}~\bibnamefont {Betto}}, \bibinfo {author}
  {\bibfnamefont {E.}~\bibnamefont {Velez-Fort}}, \bibinfo {author}
  {\bibfnamefont {A.}~\bibnamefont {Amorese}}, \bibinfo {author} {\bibfnamefont
  {G.}~\bibnamefont {Ghiringhelli}}, \bibinfo {author} {\bibfnamefont
  {L.}~\bibnamefont {Braicovich}}, \bibinfo {author} {\bibfnamefont
  {R.}~\bibnamefont {Barrett}}, \bibinfo {author} {\bibfnamefont
  {G.}~\bibnamefont {Berruyer}}, \bibinfo {author} {\bibfnamefont
  {F.}~\bibnamefont {Cianciosi}}, \bibinfo {author} {\bibfnamefont
  {L.}~\bibnamefont {Eybert}}, \bibinfo {author} {\bibfnamefont
  {P.}~\bibnamefont {Marion}}, \bibinfo {author} {\bibfnamefont
  {P.}~\bibnamefont {van~der Linden}}, \ and\ \bibinfo {author} {\bibfnamefont
  {L.}~\bibnamefont {Zhang}},\ }\href {\doibase
  https://doi.org/10.1016/j.nima.2018.07.001} {\bibfield  {journal} {\bibinfo
  {journal} {Nucl. Instrum. Methods A}\ }\textbf {\bibinfo {volume} {903}},\
  \bibinfo {pages} {175} (\bibinfo {year} {2018})}\BibitemShut {NoStop}%
\bibitem [{\citenamefont {Afonso}\ and\ \citenamefont
  {Kune\v{s}}(2017)}]{Afonso2017prb}%
  \BibitemOpen
  \bibfield  {author} {\bibinfo {author} {\bibfnamefont {J.~F.}\ \bibnamefont
  {Afonso}}\ and\ \bibinfo {author} {\bibfnamefont {J.}~\bibnamefont
  {Kune\v{s}}},\ }\href {\doibase 10.1103/PhysRevB.95.115131} {\bibfield
  {journal} {\bibinfo  {journal} {Phys. Rev. B}\ }\textbf {\bibinfo {volume}
  {95}},\ \bibinfo {pages} {115131} (\bibinfo {year} {2017})}\BibitemShut
  {NoStop}%
\bibitem [{\citenamefont {Georges}\ \emph {et~al.}(1996)\citenamefont
  {Georges}, \citenamefont {Kotliar}, \citenamefont {Krauth},\ and\
  \citenamefont {Rozenberg}}]{Georges96}%
  \BibitemOpen
  \bibfield  {author} {\bibinfo {author} {\bibfnamefont {A.}~\bibnamefont
  {Georges}}, \bibinfo {author} {\bibfnamefont {G.}~\bibnamefont {Kotliar}},
  \bibinfo {author} {\bibfnamefont {W.}~\bibnamefont {Krauth}}, \ and\ \bibinfo
  {author} {\bibfnamefont {M.~J.}\ \bibnamefont {Rozenberg}},\ }\href {\doibase
  10.1103/RevModPhys.68.13} {\bibfield  {journal} {\bibinfo  {journal} {Rev.
  Mod. Phys.}\ }\textbf {\bibinfo {volume} {68}},\ \bibinfo {pages} {13}
  (\bibinfo {year} {1996})}\BibitemShut {NoStop}%
\bibitem [{\citenamefont {Byczuk}\ and\ \citenamefont
  {Vollhardt}(2008)}]{Byczuk08}%
  \BibitemOpen
  \bibfield  {author} {\bibinfo {author} {\bibfnamefont {K.}~\bibnamefont
  {Byczuk}}\ and\ \bibinfo {author} {\bibfnamefont {D.}~\bibnamefont
  {Vollhardt}},\ }\href {\doibase 10.1103/PhysRevB.77.235106} {\bibfield
  {journal} {\bibinfo  {journal} {Phys. Rev. B}\ }\textbf {\bibinfo {volume}
  {77}},\ \bibinfo {pages} {235106} (\bibinfo {year} {2008})}\BibitemShut
  {NoStop}%
\bibitem [{\citenamefont {Anders}\ \emph {et~al.}(2010)\citenamefont {Anders},
  \citenamefont {Gull}, \citenamefont {Pollet}, \citenamefont {Troyer},\ and\
  \citenamefont {Werner}}]{Anders10}%
  \BibitemOpen
  \bibfield  {author} {\bibinfo {author} {\bibfnamefont {P.}~\bibnamefont
  {Anders}}, \bibinfo {author} {\bibfnamefont {E.}~\bibnamefont {Gull}},
  \bibinfo {author} {\bibfnamefont {L.}~\bibnamefont {Pollet}}, \bibinfo
  {author} {\bibfnamefont {M.}~\bibnamefont {Troyer}}, \ and\ \bibinfo {author}
  {\bibfnamefont {P.}~\bibnamefont {Werner}},\ }\href {\doibase
  10.1103/PhysRevLett.105.096402} {\bibfield  {journal} {\bibinfo  {journal}
  {Phys. Rev. Lett.}\ }\textbf {\bibinfo {volume} {105}},\ \bibinfo {pages}
  {096402} (\bibinfo {year} {2010})}\BibitemShut {NoStop}%
\bibitem [{\citenamefont {Anders}\ \emph {et~al.}(2011)\citenamefont {Anders},
  \citenamefont {Gull}, \citenamefont {Pollet}, \citenamefont {Troyer},\ and\
  \citenamefont {Werner}}]{Anders11}%
  \BibitemOpen
  \bibfield  {author} {\bibinfo {author} {\bibfnamefont {P.}~\bibnamefont
  {Anders}}, \bibinfo {author} {\bibfnamefont {E.}~\bibnamefont {Gull}},
  \bibinfo {author} {\bibfnamefont {L.}~\bibnamefont {Pollet}}, \bibinfo
  {author} {\bibfnamefont {M.}~\bibnamefont {Troyer}}, \ and\ \bibinfo {author}
  {\bibfnamefont {P.}~\bibnamefont {Werner}},\ }\href {\doibase
  10.1088/1367-2630/13/7/075013} {\bibfield  {journal} {\bibinfo  {journal}
  {New J. Phys.}\ }\textbf {\bibinfo {volume} {13}},\ \bibinfo {pages} {075013}
  (\bibinfo {year} {2011})}\BibitemShut {NoStop}%
\bibitem [{\citenamefont {Hu}\ and\ \citenamefont {Tong}(2009)}]{Hu09}%
  \BibitemOpen
  \bibfield  {author} {\bibinfo {author} {\bibfnamefont {W.-J.}\ \bibnamefont
  {Hu}}\ and\ \bibinfo {author} {\bibfnamefont {N.-H.}\ \bibnamefont {Tong}},\
  }\href {\doibase 10.1103/PhysRevB.80.245110} {\bibfield  {journal} {\bibinfo
  {journal} {Phys. Rev. B}\ }\textbf {\bibinfo {volume} {80}},\ \bibinfo
  {pages} {245110} (\bibinfo {year} {2009})}\BibitemShut {NoStop}%
\bibitem [{\citenamefont {Werner}\ \emph {et~al.}(2006)\citenamefont {Werner},
  \citenamefont {Comanac}, \citenamefont {de' Medici}, \citenamefont {Troyer},\
  and\ \citenamefont {Millis}}]{Werner06}%
  \BibitemOpen
  \bibfield  {author} {\bibinfo {author} {\bibfnamefont {P.}~\bibnamefont
  {Werner}}, \bibinfo {author} {\bibfnamefont {A.}~\bibnamefont {Comanac}},
  \bibinfo {author} {\bibfnamefont {L.}~\bibnamefont {de' Medici}}, \bibinfo
  {author} {\bibfnamefont {M.}~\bibnamefont {Troyer}}, \ and\ \bibinfo {author}
  {\bibfnamefont {A.~J.}\ \bibnamefont {Millis}},\ }\href {\doibase
  10.1103/PhysRevLett.97.076405} {\bibfield  {journal} {\bibinfo  {journal}
  {Phys. Rev. Lett.}\ }\textbf {\bibinfo {volume} {97}},\ \bibinfo {pages}
  {076405} (\bibinfo {year} {2006})}\BibitemShut {NoStop}%
\bibitem [{Note1()}]{Note1}%
  \BibitemOpen
  \bibinfo {note} {We have checked on several cases that both approaches lead
  to the same low-energy spectral functions and excitation concentrations, see
  SM~\cite {sm}.}\BibitemShut {Stop}%
\bibitem [{\citenamefont {Shastry}(2011)}]{Shastry11}%
  \BibitemOpen
  \bibfield  {author} {\bibinfo {author} {\bibfnamefont {B.~S.}\ \bibnamefont
  {Shastry}},\ }\href {\doibase 10.1103/PhysRevLett.107.056403} {\bibfield
  {journal} {\bibinfo  {journal} {Phys. Rev. Lett.}\ }\textbf {\bibinfo
  {volume} {107}},\ \bibinfo {pages} {056403} (\bibinfo {year}
  {2011})}\BibitemShut {NoStop}%
\bibitem [{\citenamefont {Perepelitsky}\ and\ \citenamefont
  {Shastry}(2013)}]{Perepelitsky13}%
  \BibitemOpen
  \bibfield  {author} {\bibinfo {author} {\bibfnamefont {E.}~\bibnamefont
  {Perepelitsky}}\ and\ \bibinfo {author} {\bibfnamefont {B.~S.}\ \bibnamefont
  {Shastry}},\ }\href {\doibase https://doi.org/10.1016/j.aop.2013.09.010}
  {\bibfield  {journal} {\bibinfo  {journal} {Annals of Physics}\ }\textbf
  {\bibinfo {volume} {338}},\ \bibinfo {pages} {283 } (\bibinfo {year}
  {2013})}\BibitemShut {NoStop}%
\bibitem [{\citenamefont {Fern\'andez~Afonso}\ \emph
  {et~al.}(2018)\citenamefont {Fern\'andez~Afonso}, \citenamefont {Sotnikov},\
  and\ \citenamefont {Kune\v{s}}}]{Afonso2018jccm}%
  \BibitemOpen
  \bibfield  {author} {\bibinfo {author} {\bibfnamefont {J.}~\bibnamefont
  {Fern\'andez~Afonso}}, \bibinfo {author} {\bibfnamefont {A.}~\bibnamefont
  {Sotnikov}}, \ and\ \bibinfo {author} {\bibfnamefont {J.}~\bibnamefont
  {Kune\v{s}}},\ }\href {http://stacks.iop.org/0953-8984/30/i=13/a=135603}
  {\bibfield  {journal} {\bibinfo  {journal} {J. Phys.: Condens. Matter}\
  }\textbf {\bibinfo {volume} {30}},\ \bibinfo {pages} {135603} (\bibinfo
  {year} {2018})}\BibitemShut {NoStop}%
\bibitem [{\citenamefont {Zhang}\ \emph
  {et~al.}(2012{\natexlab{b}})\citenamefont {Zhang}, \citenamefont {Gorelov},
  \citenamefont {Koch},\ and\ \citenamefont {Pavarini}}]{Zhang2012}%
  \BibitemOpen
  \bibfield  {author} {\bibinfo {author} {\bibfnamefont {G.}~\bibnamefont
  {Zhang}}, \bibinfo {author} {\bibfnamefont {E.}~\bibnamefont {Gorelov}},
  \bibinfo {author} {\bibfnamefont {E.}~\bibnamefont {Koch}}, \ and\ \bibinfo
  {author} {\bibfnamefont {E.}~\bibnamefont {Pavarini}},\ }\href {\doibase
  10.1103/PhysRevB.86.184413} {\bibfield  {journal} {\bibinfo  {journal} {Phys.
  Rev. B}\ }\textbf {\bibinfo {volume} {86}},\ \bibinfo {pages} {184413}
  (\bibinfo {year} {2012}{\natexlab{b}})}\BibitemShut {NoStop}%
\bibitem [{\citenamefont {K\v{r}\'apek}\ \emph {et~al.}(2012)\citenamefont
  {K\v{r}\'apek}, \citenamefont {Nov\'ak}, \citenamefont {Kune\v{s}},
  \citenamefont {Novoselov}, \citenamefont {Korotin},\ and\ \citenamefont
  {Anisimov}}]{Krapek12}%
  \BibitemOpen
  \bibfield  {author} {\bibinfo {author} {\bibfnamefont {V.}~\bibnamefont
  {K\v{r}\'apek}}, \bibinfo {author} {\bibfnamefont {P.}~\bibnamefont
  {Nov\'ak}}, \bibinfo {author} {\bibfnamefont {J.}~\bibnamefont {Kune\v{s}}},
  \bibinfo {author} {\bibfnamefont {D.}~\bibnamefont {Novoselov}}, \bibinfo
  {author} {\bibfnamefont {D.~M.}\ \bibnamefont {Korotin}}, \ and\ \bibinfo
  {author} {\bibfnamefont {V.~I.}\ \bibnamefont {Anisimov}},\ }\href {\doibase
  10.1103/PhysRevB.86.195104} {\bibfield  {journal} {\bibinfo  {journal} {Phys.
  Rev. B}\ }\textbf {\bibinfo {volume} {86}},\ \bibinfo {pages} {195104}
  (\bibinfo {year} {2012})}\BibitemShut {NoStop}%
\bibitem [{\citenamefont {Tomiyasu}\ \emph {et~al.}()\citenamefont {Tomiyasu},
  \citenamefont {Nomura}, \citenamefont {Kobayashi}, \citenamefont {Ishihara},
  \citenamefont {Ohira-Kawamura},\ and\ \citenamefont
  {Kofu}}]{Tomiyasu2018arXiv}%
  \BibitemOpen
  \bibfield  {author} {\bibinfo {author} {\bibfnamefont {K.}~\bibnamefont
  {Tomiyasu}}, \bibinfo {author} {\bibfnamefont {T.}~\bibnamefont {Nomura}},
  \bibinfo {author} {\bibfnamefont {Y.}~\bibnamefont {Kobayashi}}, \bibinfo
  {author} {\bibfnamefont {S.}~\bibnamefont {Ishihara}}, \bibinfo {author}
  {\bibfnamefont {S.}~\bibnamefont {Ohira-Kawamura}}, \ and\ \bibinfo {author}
  {\bibfnamefont {M.}~\bibnamefont {Kofu}},\ }\href@noop {} {\ }\Eprint
  {http://arxiv.org/abs/1808.05888} {arXiv:1808.05888} \BibitemShut {NoStop}%
\bibitem [{\citenamefont {Kune\v{s}}\ \emph {et~al.}(2010)\citenamefont
  {Kune\v{s}}, \citenamefont {Arita}, \citenamefont {Wissgott}, \citenamefont
  {Toschi}, \citenamefont {Ikeda},\ and\ \citenamefont {Held}}]{wien2wannier}%
  \BibitemOpen
  \bibfield  {author} {\bibinfo {author} {\bibfnamefont {J.}~\bibnamefont
  {Kune\v{s}}}, \bibinfo {author} {\bibfnamefont {R.}~\bibnamefont {Arita}},
  \bibinfo {author} {\bibfnamefont {P.}~\bibnamefont {Wissgott}}, \bibinfo
  {author} {\bibfnamefont {A.}~\bibnamefont {Toschi}}, \bibinfo {author}
  {\bibfnamefont {H.}~\bibnamefont {Ikeda}}, \ and\ \bibinfo {author}
  {\bibfnamefont {K.}~\bibnamefont {Held}},\ }\href {\doibase
  http://dx.doi.org/10.1016/j.cpc.2010.08.005} {\bibfield  {journal} {\bibinfo
  {journal} {Comput. Phys. Commun.}\ }\textbf {\bibinfo {volume} {181}},\
  \bibinfo {pages} {1888} (\bibinfo {year} {2010})}\BibitemShut {NoStop}%
\bibitem [{\citenamefont {Mostofi}\ \emph {et~al.}(2014)\citenamefont
  {Mostofi}, \citenamefont {Yates}, \citenamefont {Pizzi}, \citenamefont {Lee},
  \citenamefont {S.}, \citenamefont {Vanderbilt},\ and\ \citenamefont
  {Marzari}}]{wannier90}%
  \BibitemOpen
  \bibfield  {author} {\bibinfo {author} {\bibfnamefont {A.~A.}\ \bibnamefont
  {Mostofi}}, \bibinfo {author} {\bibfnamefont {J.~R.}\ \bibnamefont {Yates}},
  \bibinfo {author} {\bibfnamefont {G.}~\bibnamefont {Pizzi}}, \bibinfo
  {author} {\bibfnamefont {Y.-S.}\ \bibnamefont {Lee}}, \bibinfo {author}
  {\bibfnamefont {I.}~\bibnamefont {S.}}, \bibinfo {author} {\bibfnamefont
  {D.}~\bibnamefont {Vanderbilt}}, \ and\ \bibinfo {author} {\bibfnamefont
  {N.}~\bibnamefont {Marzari}},\ }\href {\doibase
  http://dx.doi.org/10.1016/j.cpc.2014.05.003} {\bibfield  {journal} {\bibinfo
  {journal} {Comput. Phys. Commun.}\ }\textbf {\bibinfo {volume} {185}},\
  \bibinfo {pages} {2309} (\bibinfo {year} {2014})}\BibitemShut {NoStop}%
\bibitem [{\citenamefont {Ikeda}\ \emph {et~al.}(2016)\citenamefont {Ikeda},
  \citenamefont {Nomura}, \citenamefont {Matsuda}, \citenamefont {Matsuo},
  \citenamefont {Kindo},\ and\ \citenamefont {Sato}}]{Ikeda2016prb}%
  \BibitemOpen
  \bibfield  {author} {\bibinfo {author} {\bibfnamefont {A.}~\bibnamefont
  {Ikeda}}, \bibinfo {author} {\bibfnamefont {T.}~\bibnamefont {Nomura}},
  \bibinfo {author} {\bibfnamefont {Y.~H.}\ \bibnamefont {Matsuda}}, \bibinfo
  {author} {\bibfnamefont {A.}~\bibnamefont {Matsuo}}, \bibinfo {author}
  {\bibfnamefont {K.}~\bibnamefont {Kindo}}, \ and\ \bibinfo {author}
  {\bibfnamefont {K.}~\bibnamefont {Sato}},\ }\href {\doibase
  10.1103/PhysRevB.93.220401} {\bibfield  {journal} {\bibinfo  {journal} {Phys.
  Rev. B}\ }\textbf {\bibinfo {volume} {93}},\ \bibinfo {pages} {220401}
  (\bibinfo {year} {2016})}\BibitemShut {NoStop}%
\bibitem [{\citenamefont {Afonso}\ \emph {et~al.}(2019)\citenamefont {Afonso},
  \citenamefont {Sotnikov}, \citenamefont {Hariki},\ and\ \citenamefont
  {Kune\ifmmode~\check{s}\else \v{s}\fi{}}}]{Afonso19}%
  \BibitemOpen
  \bibfield  {author} {\bibinfo {author} {\bibfnamefont {J.~F.}\ \bibnamefont
  {Afonso}}, \bibinfo {author} {\bibfnamefont {A.}~\bibnamefont {Sotnikov}},
  \bibinfo {author} {\bibfnamefont {A.}~\bibnamefont {Hariki}}, \ and\ \bibinfo
  {author} {\bibfnamefont {J.}~\bibnamefont {Kune\ifmmode~\check{s}\else
  \v{s}\fi{}}},\ }\href {\doibase 10.1103/PhysRevB.99.205118} {\bibfield
  {journal} {\bibinfo  {journal} {Phys. Rev. B}\ }\textbf {\bibinfo {volume}
  {99}},\ \bibinfo {pages} {205118} (\bibinfo {year} {2019})}\BibitemShut
  {NoStop}%
\bibitem [{\citenamefont {Jarrell}\ and\ \citenamefont
  {Gubernatis}(1996)}]{jarrell96}%
  \BibitemOpen
  \bibfield  {author} {\bibinfo {author} {\bibfnamefont {M.}~\bibnamefont
  {Jarrell}}\ and\ \bibinfo {author} {\bibfnamefont {J.}~\bibnamefont
  {Gubernatis}},\ }\href {\doibase
  http://dx.doi.org/10.1016/0370-1573(95)00074-7} {\bibfield  {journal}
  {\bibinfo  {journal} {Phys. Rep.}\ }\textbf {\bibinfo {volume} {269}},\
  \bibinfo {pages} {133 } (\bibinfo {year} {1996})}\BibitemShut {NoStop}%
\bibitem [{\citenamefont {Gull}\ \emph {et~al.}(2011)\citenamefont {Gull},
  \citenamefont {Millis}, \citenamefont {Lichtenstein}, \citenamefont
  {Rubtsov}, \citenamefont {Troyer},\ and\ \citenamefont {Werner}}]{Gull11}%
  \BibitemOpen
  \bibfield  {author} {\bibinfo {author} {\bibfnamefont {E.}~\bibnamefont
  {Gull}}, \bibinfo {author} {\bibfnamefont {A.~J.}\ \bibnamefont {Millis}},
  \bibinfo {author} {\bibfnamefont {A.~I.}\ \bibnamefont {Lichtenstein}},
  \bibinfo {author} {\bibfnamefont {A.~N.}\ \bibnamefont {Rubtsov}}, \bibinfo
  {author} {\bibfnamefont {M.}~\bibnamefont {Troyer}}, \ and\ \bibinfo {author}
  {\bibfnamefont {P.}~\bibnamefont {Werner}},\ }\href {\doibase
  10.1103/RevModPhys.83.349} {\bibfield  {journal} {\bibinfo  {journal} {Rev.
  Mod. Phys.}\ }\textbf {\bibinfo {volume} {83}},\ \bibinfo {pages} {349}
  (\bibinfo {year} {2011})}\BibitemShut {NoStop}%
\end{thebibliography}%
	
\end{document}